\documentclass[12pt]{JHEP3}
\usepackage{amsmath,amssymb,epsfig}

\newcommand{\be}{\begin{equation}}
\newcommand{\ee}{\end{equation}}
\newcommand{\beqa}{\begin{eqnarray}}
\newcommand{\eeqa}{\end{eqnarray}}
 \newcommand{\bn}{\begin{enumerate}}
\newcommand{\en}{\end{enumerate}}

\def\bra#1{\left\langle #1\right|}
\def\eeq{\end{equation}}

\def\ket#1{\left| #1\right\rangle}

\def\Tr{\mathop{\rm Tr}}

\relax

\title{Robustness and Infrared Sensitivity of 
Various Observables in the Application 
of AdS/CFT to Heavy Ion Collisions}

\author{Hong Liu, Krishna Rajagopal and Yeming Shi\\

\vspace{0.2in}

Center for Theoretical Physics, \\
  Massachusetts
Institute of Technology, \\
Cambridge, MA 02139, USA \vspace{0.1in}

E-mail addresses: {\tt hong\_liu@mit.edu, krishna@ctp.mit.edu,
yeming@mit.edu} 
}

\abstract{
We investigate the robustness with respect to the introduction
of nonconformality of five properties of strongly
coupled plasmas that have been calculated in ${\cal N}=4$ supersymmetric
Yang-Mills (SYM) theory at nonzero temperature, motivated by
the goal of understanding phenomena in relativistic heavy ion collisions.
(The five properties are the jet quenching parameter, the
velocity dependence of screening, and the drag and transverse
and longitudinal momentum diffusion coefficients for a heavy quark
pulled through the plasma.)  We do so using a toy model
in which nonconformality is introduced via a one-parameter
deformation of the AdS black hole dual to the hot ${\cal N}=4$ SYM plasma.
For values of this parameter which
correspond to a degree of nonconformality comparable
to that seen in lattice calculations of
QCD thermodynamics at temperatures a few
times that of the crossover to quark-gluon plasma,
we find that the jet quenching parameter is affected by the
nonconformality at the 30\% level or less, the screening
length is affected at the 20\% level or less, but
the drag and diffusion coefficients for a slowly moving heavy quark
can be modified by 
as much as 80\%.  
However,
we show that all but one of the five properties that we investigate become
completely insensitive to the nonconformality in the high
velocity limit $v\rightarrow 1$. The exception is the 
jet quenching parameter,
which is unique among the quantities that we investigate
in being ``infrared sensitive'' even at $v=1$, where it is
defined.  That is, it is the only high-velocity observable
that we investigate which is sensitive to properties of
the medium at infrared energy scales proportional to $T$, 
namely the scales where the quark-gluon plasma of QCD can
be strongly coupled.  The other four quantities all probe 
only scales that are larger than $T$ by a factor that
diverges as $v\rightarrow 1$, namely scales where the 
${\cal N}=4$ SYM plasma can be strongly coupled but the 
quark-gluon plasma of QCD is not. 
}

\preprint{MIT-CTP-3932}

\begin{document}
\def\vev#1{\langle#1\rangle}
\def\ov{\over}
\def\le{\left}
\def\ri{\right}
\def\ha{{1\over 2}}
\def\lam{{\lambda}}
\def\Lam{{\Lambda}}
\def\al{{\alpha}}
\def\ket#1{|#1\rangle}
\def\bra#1{\langle#1|}
\def\vev#1{\langle#1\rangle}
\def\det{{\rm det}}
\def\tr{{\rm tr}}
\def\Tr{{\rm Tr}}
\def\NN{{\cal N}}
\def\th{{\theta}}

\def\Om{{\Omega}}
\def \th{{\theta}}

\def \lam {\lambda}
\def \om {\omega}
\def \ra {\rightarrow}
\def \ga {\gamma}
\def\sig{{\sigma}}
\def\ep{{\epsilon}}
\def\apr{{\alpha'}}
\newcommand{\p}{\partial}
\def\LL{{\cal L}}
\def\HH{{\cal H}}
\def\GG{{\cal G}}
\def\TT{{\cal T}}
\def\CC{{\cal C}}
\def\OO{{\cal O}}
\def\PP{{\cal P}}
\def\tir{{\tilde r}}

\newcommand{\bea}{\begin{eqnarray}}
\newcommand{\eea}{\end{eqnarray}}
\newcommand{\nn}{\nonumber\\}

\section{Introduction and Summary}

The AdS/CFT correspondence~\cite{AdS/CFT} has provided an
important tool for understanding the dynamics of 
many and varied strongly
coupled gauge theories.  By now, it has been applied at nonzero
temperature to gauge theory plasmas in theories that are conformal,
or not; theories that are confining at zero temperature, or not; theories
with varying degrees of supersymmetry; theories which
at weak coupling contain both fundamentals and adjoints, or only
adjoints; to plasmas with zero or nonzero chemical
potentials; to plasmas that are static or
expanding.    In terms that are qualitative enough to apply to all
these examples, the correspondence states that
a (3+1)-dimensional gauge theory plasma 
at some temperature $T$ is equivalent to a (particular) string theory
in a (particular) curved higher-dimensional spacetime which includes a 
black hole horizon with Hawking temperature $T$.  In the limit
in which $N_c$, the number of colors in the gauge theory,
and $\lambda\equiv g^2 N_c$, the 't Hooft coupling of the gauge
theory, are both taken to infinity, the equivalent (dual)
gravity description of the strongly coupled
gauge theory plasma becomes classical.  This means that, in the
regime of large $N_c$ and strong coupling, calculations of
various dynamical properties of strongly coupled gauge theory
plasmas (that are difficult to calculate in the gauge theory {\it per se})
become equivalent to tractable calculations in a 
classical spacetime background.  We shall specify five examples of such
calculations below.  

The simplest, most symmetric, example of a gauge theory
whose dual gravity description is the original example discussed
by Maldacena: ${\cal N}=4$ supersymmetric Yang-Mills theory (SYM), which
at nonzero temperature is dual to 
Type IIB string theory a $(9+1)$-dimensional
spacetime given by $(4+1)$-dimensional Anti-de Sitter (AdS)
space, with the five remaining compact dimensions forming an $S_5$.  
The metric for the AdS black hole can be written as
\begin{align}
ds^{2} & =
 -\frac{r^{2}}{R^{2}}
 \le(1-\frac{r_{0}^{4}}{r^{4}}\ri)dt^{2}
 +\frac{r^{2}}{R^{2}}d\vec{x}^{2}+\frac{R^{2}}{r^{2}}\frac{dr^{2}}{1-\frac{r_{0}^{4}}{r^{4}}}
\nonumber\\
&=
\frac{R^{2}}{z^2}
\le[-\le(1-\frac{z^{4}}{z_{0}^4}\ri)dt^{2}+dx_{1}^{2}+dx_{2}^{2}+dx_{3}^{2}
+\frac{dz^{2}}{1-\frac{z^{4}}{z_{0}^4}}\ri]\ ,\label{eq:AdSmetric}
\end{align}
where $R$ is the AdS curvature, 
where $z=R^2/r$, and where the black hole horizon is at $r=r_0=\pi R^2 T$,
meaning $z=z_0=\frac{1}{\pi T}$.  In
some respects, the gauge theory can be thought
of as living at the $(3+1)$-dimensional ``boundary'' 
$z=0$. However, it is important to remember
that the equivalence between the gauge theory and its gravity 
description is holographic, in that all of the physics at varying values of $z$
in the gravity description is encoded in the gauge theory, with 
the fifth-dimension-position $z$ in the spacetime  (\ref{eq:AdSmetric})
corresponding to  length scale in the $(3+1)$-dimensional gauge 
theory~\cite{AdS/CFT,IRUV}.

Although many (in fact infinitely many)
other examples of gauge theories with dual gravity descriptions are 
known, such a description has not yet been found for
$SU(N_c)$ gauge theory (with
or without quarks in the fundamental representation). 
And, furthermore, all known theories with gravity
duals differ from QCD in important respects.  
Taking
${\cal N}=4$ SYM as an example,
at weak coupling it has more
adjoint degrees of freedom 
than in QCD,
it has no fundamental degrees
of freedom, and it is conformal.  
And, at zero temperature ${\cal N}=4$ SYM  is supersymmetric and
does not
feature either confinement or chiral symmetry breaking.
However, the plasmas of the two theories,
namely ${\cal N}=4$ SYM at $T>0$ and QCD at $T$ 
above $T_c\sim 170$ MeV, are more similar than their
vacua.    Neither plasma confines or breaks chiral symmetry, and
neither is supersymmetric since $T\neq 0$ breaks supersymmetry.
The successful comparison of data from heavy ion collisions
at RHIC (on azimuthally asymmetric collective flow) with ideal 
(zero shear viscosity $\eta$)
hydrodynamics indicates that, somewhat above $T_c$, the QCD plasma
is a strongly coupled liquid~\cite{RHIC}.   
Strongly coupled liquids may not have
any well-defined quasiparticles, so the differences between the
quasiparticles of the two theories at weak coupling need not be important,
at least for judiciously chosen ratios of observables.  And, lattice 
calculations~\cite{Boyd:1996bx,
Cheng:2007jq,Fodor:2007sy,Karsch:2006xs}
indicate that above $\sim 2 T_c$, the thermodynamics of the QCD plasma 
becomes approximately scale invariant.
More generally speaking, it is often the case that
macroscopic phenomena in a sufficiently excited many-body system
are common across large universality classes of theories that differ in
many (microscopic) respects.  This raises the exciting possibility
that one may be able to gain insights into the
thermodynamics and dynamics of the strongly coupled
plasma of QCD using calculations in other gauge theories whose
gravity duals are currently known.

Many authors have developed the strategy
of calculating dynamical properties of gauge theory
plasmas (that are of interest because they can be related 
 to phenomena in heavy ion collision experiments) by calculating
 them in ${\cal N}=4$ SYM and other theories with gravity duals.
 Turning the qualitative  insights obtained in this way into
semiquantitative inferences for QCD (or even for QCD at large $N_c$)
requires understanding what observables are universal across
what classes of strongly coupled plasmas or, if not that, understanding
how observables change as the strongly coupled ${\cal N}=4$ SYM
plasma is deformed in various ways that make it more QCD-like.
At present, the quantity for which the evidence of a universality
of this sort is strongest is $\eta/s$, the ratio of the 
shear viscosity to the entropy density: in the large $N_c$ and
strong coupling limit, 
it is given by $1/4\pi$ for
{\it any} gauge theory with a dual gravity 
description~\cite{Policastro:2001yc,Kovtun:2003wp,Buchel:2003tz}.
The discovery that 
in an infinite class of conformal gauge theories
the jet quenching parameter $\hat q$,
that we shall discuss below, is given by $\sqrt{\lambda} T^3$
times a pure number
that is proportional to $\sqrt{s}/N_c$ suggests
a second quantity with a degree
of universality~\cite{Liu:2006he}, but 
one that at at present is only known 
to apply to conformal theories.

It is clearly critical to extend AdS/CFT calculations 
of dynamical properties of gauge theory plasmas
to nonconformal theories.
Unfortunately, the known examples of nonconformal gauge
theories with gravity duals are rather complicated
at nonzero temperature, see for example Refs.~\cite{Buchel:2001gw,Buchel:2003ah}, making
it hard to extract insights from them without extensive, probably
numerical, study.  
Here we will take a pragmatic approach, using a simple
toy model, similar to that introduced at zero temperature 
in Ref.~\cite{Karch:2006pv} and at nonzero temperature in
Refs.~\cite{Herzog:2006ra,Nakano:2006js}
and in the form that we shall use by Kajantie, 
Tahkokallio and Yee~\cite{Kajantie:2006hv}, in which (\ref{eq:AdSmetric}) is
deformed into the string frame metric
\begin{align}
ds^{2} & =\frac{R^{2}e^{\frac{29cz^{2}}{20}}}{z^{2}}
\le[-\le(1-\frac{z^{4}}{z_{0}^4}\ri)dt^{2}+dx_{1}^{2}+dx_{2}^{2}+dx_{3}^{2}
+\frac{dz^{2}}{1-\frac{z^{4}}{z_{0}^4}}\ri] \nonumber\\
 & =e^{\frac{29}{20}c\frac{R^{4}}{r^{2}}}\le[-\frac{r^{2}}{R^{2}}
 \le(1-\frac{r_{0}^{4}}{r^{4}}\ri)dt^{2}
 +\frac{r^{2}}{R^{2}}d\vec{x}^{2}+\frac{R^{2}}{r^{2}}\frac{dr^{2}}{1-\frac{r_{0}^{4}}{r^{4}}}
 \ri]\ . \label{eq:fmetric}
\end{align}
Here, the dimensionful quantity $c$ defines a one-parameter nonconformal deformation
of the AdS black hole.  Certainly our investigations should also be repeated
for other examples of such deformations.  The advantage of using the specific
form (\ref{eq:fmetric}) is its tractability together with the fact
that the authors of Ref.~\cite{Kajantie:2006hv} have estimated that choosing
$c\simeq 0.127$~GeV$^2$ makes the thermodynamics of this toy model most similar  
to QCD thermodynamics, determined by lattice calculations. Specifically, they
introduce a second toy model for QCD below $T_c$, choose its parameters
to give a reasonable meson spectrum in vacuum, and then find that $c=0.127$~GeV$^2$ puts
the transition between their low and high temperature models --- whose construction
is their purpose --- at $T_c=170$~MeV, as 
in QCD.  

We shall determine
how five dynamical observables, previously calculated at $c=0$, depend on $c$.
Since in the absence of $c$ the only dimensionful quantity in the 
otherwise conformal theory
is $T$, the magnitude of the nonconformal effects
that we compute must be controlled by the dimensionless ratio $c/T^2$.  
We shall plot our results for values of this parameter that lie within the
range $0\leq c/T^2 \leq 4$,
which
corresponds to allowing a $c$ as large as 0.36~GeV$^2$
at $T=300$~MeV.
Note that the metric~(\ref{eq:fmetric}) does not correspond to
a solution to supergravity equations of motion, and note furthermore that
the form of the metric for the five compact dimensions is unspecified.  These ambiguities
are what make the model a model: with $c\neq 0$,  it is impossible to say what, if any,
gauge theory the metric (\ref{eq:fmetric}) is dual to.  This makes it impossible
to give a rigorous determination of its entropy density $s$, as the authors 
of Ref.~\cite{Kajantie:2006hv}
explain, or to determine its weak coupling degrees of freedom.   So, we shall not
use this model to test how other observables depend on these quantities.
Our sole
purpose is to explore the effects of the introduction of nonconformality.

Although
it is not possible to give a rigorous argument 
for the entropy density $s$ corresponding to the metric
(\ref{eq:fmetric}),
given that the metric is not known to be a solution to supergravity
equations of motion, the authors of Ref.~\cite{Kajantie:2006hv} have conjectured
that  $s$ is given by
\be
s=\frac{\pi^2 N_c^2 T^3}{2}\exp\left(-\frac{3}{2\pi^2}\frac{c}{T^2}\right)\ .
\ee
We can use this expression to estimate the range of values of $c$ that
compare reasonably to QCD thermodynamics, as follows.
We take this expression and obtain
the energy density $\varepsilon$ from $d\varepsilon/dT=T ds/dT$, the pressure 
$P=Ts-\varepsilon$, and
then $(\varepsilon - 3 P)/\varepsilon$ which is a measure of nonconformality.
We then find that fitting the results for this quantity in the toy model
we are using to the lattice calculations of this quantity in QCD from
Ref.~\cite{Cheng:2007jq} requires $c$ varying from $c\approx 0.18$~GeV$^2$ 
at $T=300$~MeV to $c\approx 0.11$~GeV$^2$ at $T=700$~MeV,
and $c\simeq 0.13$~GeV$^2$ does reasonably well over this
entire temperature range.
This gives 
us further confidence  that when we plot our results over the 
range $0\leq c/T^2\leq 4$ we are turning on a degree of nonconformality
that encompasses and exceeds that observed 
in QCD thermodynamics at $T=300$~MeV. 
At this temperature, the range $0.11$~GeV$^2 <c<0.18$~GeV$^2$
corresponds to $1.2 < c/T^2 < 2.0$.   Keep in mind that although $c$
is the fixed parameter in
the model, it will enter all of our results  through the
dimensionless parameter $c/T^2$. So, when we plot our results
over $0<c/T^2<4$, we can think of the higher (lower) 
values of  $c/T^2$ as corresponding to
lower (higher) temperatures.

We shall calculate five quantities that have previously been argued to be
of interest because, in QCD, they are related to phenomena in heavy ion collision
experiments.   We begin in Section 2 by 
calculating the jet quenching 
parameter $\hat q$, as in Refs.~\cite{Liu:2006ug,Buchel:2006bv,Liu:2006he}. 
This
property of the strongly coupled plasma enters into the description of
how a parton moving through this plasma with energy $E$ loses energy
by radiating 
gluons~\cite{Baier:1996sk,Zakharov:1997uu,Wiedemann:2000za}.
Gluon
radiation is the dominant energy loss mechanism in the limit
where $E\gg k_T \gg T$, with $k_T$ being 
the typical transverse momentum of the radiated gluons, 
and upon assuming that $\alpha_s(k_T)$ is 
small~\cite{Baier:1996sk,Zakharov:1997uu,Wiedemann:2000za,Gyulassy:2000er,jetquenchrev}.
That is, the analysis of jet quenching in this limit is based
upon the assumption that
QCD can be considered weakly coupled at the scale $k_T$, even
though its quark-gluon plasma (at scales $\sim T$) is strongly coupled.
In this regime, the gluon radiation itself 
is described via a weakly coupled
QCD formalism in which the one property of the thermal
medium that enters is $\hat q$, which must be computed at 
strong coupling.  In Section 3, we turn to probes of the plasma
in a completely
different kinematic regime.
We shall calculate three observables that describe the motion
of a heavy enough quark (mass $M$) moving through the strongly 
coupled plasma with
a low enough velocity $v$,
where the criterion that must be satisfied by $M$ and $v$
is~\cite{Gubser:2006nz,calsa,Liu:2006he}
\be
M>\frac{\sqrt{\lambda} T}{(1-v^2)^{1/4}}\ .
\label{HeavinessCriterion}
\ee
Because we are no longer taking
the $v\rightarrow 1$ limit,
even in a theory like QCD that is weakly
coupled in the ultraviolet we cannot assume that energy
loss is dominated by gluon radiation and cannot assume that
there is a separation of scales which justifies treating a part
of the problem at weak coupling even when the plasma itself
is strongly coupled. Instead, it is worth investigating a formalism
in which the entire calculation is done at strong coupling.
In the dual gravity theory, the criterion (\ref{HeavinessCriterion})
corresponds to requiring
that the velocity of the quark not exceed the local speed
of light at the position in $z$ where quarks of mass $M$ are located.
When (\ref{HeavinessCriterion}) is satisfied, the moving quark
is described in the dual gravity theory as trailing a string that drags
behind the moving quark~\cite{Herzog:2006gh,Gubser:2006bz}, 
meaning that the quark feels a drag force 
and diffuses.  
The three parameters that we calculate
are the drag coefficient $\mu$ (introduced and calculated at $c=0$
in \cite{Herzog:2006gh,Gubser:2006bz})
and the diffusion constants $\kappa_T$
and $\kappa_L$ for its transverse and longitudinal motion (introduced
and calculated 
at $c=0$ in~\cite{Casalderrey-Solana:2006rq,Gubser:2006nz,calsa}).
The effects of the nonconformal deformation of the 
AdS black hole metric on both
$\mu$ and $\hat q$ have been calculated previously 
in Ref.~\cite{Nakano:2006js}.
Finally, in Section 4 we determine how $c$ affects the velocity dependence
of the screening length $L_s$ defined by the 
potential between a quark-antiquark
pair with mass $M$ moving through the plasma with 
velocity $v$~\cite{Liu:2006nn,Peeters:2006iu,Chernicoff:2006hi,Liu:2006he},
again satisfying (\ref{HeavinessCriterion}) which in this case corresponds
to the requirement that $L_s$ be greater than the Compton wavelength
of an individual quark~\cite{Liu:2006he}.
We shall show that, for $0<c/T^2<4$, the effects of $c$ on 
the jet quenching parameter and on the screening length 
are modest.
For example, $\hat q$ increases by about 14\% (28\%) 
for $c/T^2=2$ ($c/T^2=4$) while the screening length
increases by about 9\% (20\%).  
This indicates that these quantities 
are robust against introduction of nonconformality
to a degree larger than that indicated 
by lattice study of QCD thermodynamics.  
The drag coefficient and the two momentum diffusion constants 
for a heavy slowly moving quark are somewhat
less robust, increasing by about 34\% (80\%) for $c/T^2=2$ ($c/T^2=4$).
Of course,
our conclusions are only quantititave within one toy model. 
Other examples in which
nonconformality is introduced should also be studied.

The metric (\ref{eq:fmetric}) has the feature that it becomes 
the metric (\ref{eq:AdSmetric})
of an AdS black hole
near $z=0$, but near the horizon it is modified by
the dimensionful parameter $c$.  
This allows us to address a further issue, that
is both qualitative and important.  
QCD, being asymptotically free, 
is weakly coupled in the ultraviolet.  
The plasma in a strongly coupled conformal
theory like ${\cal N}=4$ SYM is strongly
coupled in the ultraviolet as well as at scales of order $T$.
This means that the only properties of the plasma in a strongly
coupled conformal gauge theory 
that may yield insight into the strongly coupled plasma of QCD
are those properties which
are determined by the physics at  scales of order $T$, not by the ultraviolet
physics.   It is impossible to use calculations done within ${\cal N}=4$ SYM
to determine which quantities are ``infrared sensitive" in this sense, precisely
because the theory is conformal:  the parameter $z_0$ specifies the location
of the horizon and the value of
the temperature $T=1/(\pi z_0)$, namely the
gauge theory physics at scales $\sim T$, and at the same time
specifies the form of the metric (\ref{eq:AdSmetric})  at 
small $z$, namely the gauge theory
physics in the ultraviolet.   
So, seeing $z_0$ and hence $T$ occurring in the calculated
results for $\hat q$, $\mu$, 
$\kappa_T$, $\kappa_L$ and $L_s$ in ${\cal N}=4$ SYM
does not allow us to determine whether any of these quantities are infrared
sensitive.   In order to make such a determination, 
we must modify the theory in the infrared, i.e. in the vicinity of $z=z_0$,
in a way that leaves it unmodified at $z\rightarrow 0$, and determine 
which quantities are modified and which not.  Note that
in a  gauge theory 
whose gravity dual is given asymptotically (i.e. at $z\rightarrow 0$)
by the AdS black hole metric (\ref{eq:AdSmetric}), 
the parameter $z_0$ that occurs in the asymptotic metric will, in 
the generic case, not be related to
the temperature in any simple way.
Absent conformality, there is no longer any reason for the true temperature 
$T$, defined by the metric at
the horizon, to to be related
in any simple way to the parameter $z_0$ defined by the 
AdS black hole metric at $z\rightarrow 0$.\footnote{Consider the
$(4+1)$-dimensional extremal Reissner-Nordstrom black hole
as an example that is not directly relevant but in which
this disconnect is particularly dramatic:
the asymptotic metric for this
spacetime defines a $z_0$, but the Hawking temperature
is zero.} Our toy
model is not generic enough to manifest this effect --- the 
temperature remains
$1/(\pi z_0)$ even when $c\neq 0$ --- but we can nevertheless 
use the dependence
on $c/T^2$ to diagnose infrared sensitivity.

We find that $\hat q$ is infrared sensitive --- as noted above it changes
by 28\% for $c/T^2=4$.
The other four quantities that we study are 
all infrared sensitive at low velocity.  However, if
we take $v\rightarrow 1$ and $M\rightarrow \infty$ while
maintaining the criterion (\ref{HeavinessCriterion}) --- for example
by taking $M\rightarrow\infty$ first --- we find that $\mu$,
$\kappa_T$, $\kappa_L$ and $L_s$ all become infrared {\it in}sensitive.
That is, they become independent of $c/T^2$ in this limit, meaning
that they cannot see a modification of the gauge theory at
scales $\sim T$.  In this $v\rightarrow 1$ limit, they are determined entirely
by the ultraviolet physics in the gauge theory, making it unlikely
that their calculation in ${\cal N}=4$ SYM in this limit can be used
to draw quantitative lessons for QCD. The jet quenching
parameter, on the other hand, is defined at $v\equiv 1$
and is infrared sensitive. This is consistent with its
role in jet quenching calculations as the parameter through
which the physics of the strongly coupled plasma at scales
of order the temperature enters into the calculation of how
partons moving through this plasma
lose energy in the high parton energy limit.

At a qualitative level, our results for the infrared sensitivity of
all five observables can be guessed by examining how
they are computed in the strongly coupled $\NN=4$
SYM theory. The jet-quenching parameter $\hat q$
is extracted from the short-transverse-distance behavior of the
thermal expectation value of a light-like Wilson loop that
is long in light-like extent but short in transverse extent. In the dual
gravity description, this expectation value
can be calculated by finding the extremal
configuration of a string connecting a quark-anti-quark pair
moving at the speed of light. The extremal string configuration
touches the horizon~\cite{Liu:2006ug}. In the short transverse distance limit, after
subtracting the self-energy of each quark, one is left with mostly
the contribution of the 
part of the extremal string worldsheet that
is near the horizon.  It is therefore reasonable that,
upon calculation, we find that $\hat q$
is infrared sensitive, as is also expected given the role
that it plays in the theory of jet quenching.
In contrast, a heavy quark moving through the hot plasma
with velocity $v$, satisfying (\ref{HeavinessCriterion}), is 
described by the trailing string worldsheet first analyzed
in Refs.~\cite{Herzog:2006gh,Gubser:2006bz} 
which has  a ``worldsheet horizon'' on
it located at $z=z_0(1-v^2)^{1/4}$ as described in
Refs.~\cite{Gubser:2006nz,calsa}.  The quantities $\mu$, $\kappa_T$ and $\kappa_L$
are determined by the string worldsheet outside the worldsheet 
horizon, namely in the region $0<z<z_0(1-v^2)^{1/4}$.  
($\mu$ is determined by the momentum flow along the string
worldsheet outside the worldsheet horizon; the diffusion
constants $\kappa_T$ and $\kappa_L$ 
are determined from two-point functions describing
the fluctuations of the worldsheet coordinates outside
the worldsheet horizon.)
So, if we take the $v\rightarrow 1$ limit (while increasing $M$
so as to maintain (\ref{HeavinessCriterion})) we expect these
quantities to become
completely infrared {\it in}sensitive, sensitive only to the ultraviolet physics.
Our explicit calculation confirms this expectation.
The argument for the screening length is similar.  As $v\rightarrow 1$
(while maintaining (\ref{HeavinessCriterion}))
the velocity dependent screening length shrinks, 
$L_s(v)\sim L_s(0) (1-v^2)^{1/4}$~\cite{Liu:2006nn,Peeters:2006iu,Chernicoff:2006hi,Liu:2006he}, 
and the string worldsheet
bounded by the quark-antiquark pair --- which determines the
potential and hence $L_s$ --- only explores the $(4+1)$-dimensional
spacetime in the region $0<z\lesssim z_0(1-v^2)^{1/4}$. We
therefore also expect, and find, that $L_s$ is infrared insensitive
in the $v\rightarrow 1$ limit.  
It is worth noting, however, that for charmonium
(or bottomonium) mesons with velocities corresponding to
the transverse momenta with which they are produced in
RHIC (or LHC) collisions, $L_s$ remains infrared sensitive, probing
the strongly coupled medium at scales not far above $T$.
And, the
velocity-dependence of the screening length
is described reasonably well by 
$L_s(v)\sim L_s(0) (1-v^2)^{1/4}$ 
at all velocities, large or small, up
to corrections that we shall evaluate.   

So, 
the five quantities that we investigate
are robust to varying degrees, in the sense that if we turn on
nonconformality parametrized by a value of $c/T^2$
that is about twice as large as that which best approximates
QCD thermodynamics at $T=300$~MeV within the model of 
Ref.~\cite{Kajantie:2006hv}, 
the jet quenching parameter increases by about 30\%
and at low velocities the
screening length increases by about 20\% while the heavy
quark drag and momentum diffusion coefficients increase by about 80\%. 
If we then take
the limit $v\rightarrow 1$ while increasing the quark mass
$M$ so as to maintain (\ref{HeavinessCriterion}), we find
that the drag and diffusion coefficients
and the screening length 
all become completely insensitive to the nonconformal modification of
the physics at scales $\sim T$ that we have introduced. In this limit,
these quantities all become infrared insensitive.
This makes it likely that the calculation of
these quantities in a conformal theory like 
${\cal N}=4$ SYM can only be used to learn about the 
strongly coupled plasma of QCD at
a broadly qualitative level.  In contrast, the jet quenching
parameter $\hat q$ is defined at $v\equiv 1$ and is 
infrared sensitive, probing the properties of the 
plasma at scales of order the temperature where it is strongly coupled
in both QCD and ${\cal N}=4$ SYM.


\section{Jet Quenching Parameter}

The jet quenching parameter $\hat q$ is the
property of the plasma that enters into the description of
how a parton moving through this plasma with energy $E$ loses energy
by radiating gluons with typical transverse momentum $k_T$ in the limit
where $E\gg k_T \gg T$ and upon assuming that $\alpha_s(k_T)$ is small
enough that QCD can be considered weakly coupled at this scale, even
though its quark-gluon plasma (at scales $\sim T$) is strongly 
coupled~\cite{Baier:1996sk,Zakharov:1997uu,Wiedemann:2000za,Gyulassy:2000er,jetquenchrev}.
To the degree that these assumptions are valid, parton energy loss
is dominated by gluon radiation. In experiments at RHIC,
the jets studied
correspond to partons with $E$ at most a few tens of GeV~\cite{RHIC}. At the LHC, although
the quark-gluon plasma being studied is likely to be at
most a factor of two hotter
than that at RHIC, the jets whose quenching will be studied will
have energies of a few hundreds of GeV~\cite{LHC},  putting the assumptions
upon which the definition and extraction of the jet quenching parameter
is based on more quantitative footing.   

If the quark-gluon plasma were
weakly coupled, $\hat q$ would
be proportional to $\mu^2/\tilde \lambda$, where $\mu$ is the 
inverse of the Debye screening length and $\tilde\lambda$ 
is a suitably defined
mean free path for weakly coupled quasiparticles~\cite{Baier:1996sk}.  
Up to a logarithm, 
in a weakly coupled quark-gluon 
plasma $\hat q \propto g^4 N_c^2 T^3$~\cite{Baier:1996sk,Baier:2006fr}.
Wiedemann observed that, still for a weakly coupled plasma, $\hat q$ can
instead be extracted from the small-$L$ behavior of a 
rectangular adjoint Wilson loop whose long sides, of length $L^-$, 
are light-like 
and whose short sides, of length $L$, are transverse
to the light-cone~\cite{Wiedemann:2000za}.  
$L^-$ corresponds to the extent of the medium through
which the radiated gluon travels and $1/L$ corresponds to the transverse
momentum of the radiated gluon.  Wiedemann and two of us suggested
that this definition can be generalized to a strongly coupled plasma,
and calculated $\hat q$ for the strongly coupled ${\cal N}=4$ SYM plasma~\cite{Liu:2006ug}.
In this section, we repeat this calculation of $\hat q$ for the metric
(\ref{eq:fmetric}) of Ref.~\cite{Kajantie:2006hv}, deformed to introduce nonconformality.

\subsection{Calculation}

In the large $N_c$  limit, the 
expectation value of the adjoint Wilson loop is
the square of that in the fundamental representation.
If we in addition take the large $\lambda$ limit and use
the AdS/CFT correspondence, the expectation value
of the Wilson loop in the fundamental  representation
can be computed as~\cite{Rey:1998ik,Rey:1998bq}
 \be
 \vev{W(C)} = e^{ i S_I}, \qquad S_I = S(C) - 2 S_0\ ,
 \ee
where $S(C)$ is the Nambu-Goto action for the extremal worldsheet 
bounded at $z=0$ by the Wilson loop contour $C$
and $S_0$ is the 
Nambu-Goto action for an individual quark.
For a rectangular Wilson loop extending a distance $L^-$ along the
$x^-$ light-like direction and a distance $L$ along a transverse
direction, in the regime
$L^- \gg1/T \gg L$
the expectation value of the Wilson loop in the fundamental
representation takes the form~\cite{Liu:2006ug,Liu:2006he}
 \be \label{eq:qhatdef}
 \vev{W(C)} \equiv e^{-\frac{1}{8\sqrt{2}}\hat{q}L^{-}L^{2}}\ ,
 \ee
which defines the relation between the jet quenching parameter $\hat q$
and the Wilson loop.
Let us consider a more general non-conformal metric of the form
\begin{equation}
ds^{2}=g(r)\left[-(1-f(r))dt^{2}+d\vec{x}^{2}\right]+\frac{1}{h(r)}dr^{2}  ~,
\label{eq:generalmetric}
\end{equation}
which includes both (\ref{eq:AdSmetric}) and (\ref{eq:fmetric}) as special cases.
Buchel demonstrated in~\cite{Buchel:2006bv} that in the generic
spacetime metric given by (\ref{eq:generalmetric}), the extremal
string worldsheet connecting a light-like quark-antiquark pair
always touches the horizon, as had been demonstrated in Ref.~\cite{Liu:2006ug}
for the AdS black hole (\ref{eq:AdSmetric}).  And, furthermore, 
Buchel showed that upon evaluating the Wilson loop the jet
quenching parameter $\hat q$ is given in terms of the
string tension $1/(2\pi \alpha')$ and the 
functions appearing in the generic metric (\ref{eq:generalmetric}) by
\begin{equation}
\hat{q}=\frac{1}{\pi\alpha'}\le(\int_{r_{0}}^{\infty}\frac{dr}{\sqrt{fg^{3}h}}\ri)^{-1}
 \ , \label{eq:qhatresult}
\end{equation}
where $r_0$ is the coordinate of the black hole horizon.
The metric (\ref{eq:fmetric}) corresponds to
\begin{align}
g(r)&=\frac{r^{2}}{R^{2}}e^{\frac{29}{20}c\frac{R^{4}}{r^{2}}}  ~, \nonumber
 \\
f(r)&=\frac{r_{0}^{4}}{r^{4}}  ~, \\
h(r)&=\frac{r^{2}}{R^{2}}\left(1-\frac{r_{0}^{4}}{r^{4}}\right)
e^{-\frac{29}{20}c\frac{R^{4}}{r^{2}}}
 ~, \nonumber
\end{align}
and we shall assume 
that $R$ is related to $\lambda$ 
by $R^2/\alpha'=\sqrt{\lambda}$.\footnote{Since the metric (\ref{eq:fmetric}) 
reduces to the AdS black hole metric (\ref{eq:AdSmetric}) at small $z$ --- in the
ultraviolet in the field theory --- the relation between
$R$ and $\lambda$ is $R^2/\alpha'=\sqrt{\lambda}$ in the ultraviolet.
If we knew to what field theory the deformed metric (\ref{eq:fmetric}) is
dual, i.e. if we knew what the action was whose supergravity
equations of motion were solved by (\ref{eq:fmetric}), 
we can presume that $\lambda$ would run in some way.
As (\ref{eq:fmetric}) is just a toy model that we are using
to introduce nonconformality, we cannot determine how $\lambda$ runs.
So, we shall use $R^2/\alpha'=\sqrt{\lambda}$ throughout.}
Hence, we
find that in the metric~(\ref{eq:fmetric}) the jet quenching parameter is
given by
\begin{align} \label{eq:qhatfinal}
\hat{q}&=\frac{R^{4}}{\pi\alpha'}
 \le(\int_{r_{0}}^{\infty} dr \frac{
e^{-\frac{29}{20}c\frac{R^{4}}{r^{2}}}}
{r_{0}^{2}\sqrt{r^{4}-r_{0}^{4}}} \ri)^{-1}
 \nonumber \\
&=\sqrt{\lambda}\pi^{2}T^{3}
\le(\int_{1}^{\infty}dx\frac{e^{-\frac{29c}{20\pi^{2}T^{2}x^{2}}}}{\sqrt{x^{4}-1}}
\ri)^{-1}\ ,
\end{align}
where we have used 
$r_{0}=\pi R^{2}T$. 
The integral in
(\ref{eq:qhatfinal}) can be evaluated analytically, and the result
involves modified Bessel functions of the first kind~\cite{Nakano:2006js}. 
With $c=0$, it is given by $\sqrt{\pi} \Gamma(\frac{5}{4})/\Gamma(\frac{3}{4})$
which yields $\hat q$ for ${\cal N}=4$ SYM theory~\cite{Liu:2006ug}.
The result (\ref{eq:qhatfinal}) was obtained previously in 
Ref.~\cite{Nakano:2006js}.

From (\ref{eq:qhatfinal}) we see that
$\hat q \propto \lambda^{\frac{1}{2}} N_c^0$ meaning
that, with $c\neq 0$ as with $c=0$, the jet quenching parameter
is not proportional to the entropy density or to the number density of
scatterers or quasiparticles as at weak coupling~\cite{Liu:2006ug},
consistent with the absence of any quasiparticle
description of the strongly coupled plasma.
Within the formalism of Ref.~\cite{Liang:2008vz}, this qualitative conclusion 
can be phrased as the statement that multiple gluon correlations 
are just as important as two gluon correlations in the plasma
of strongly coupled ${\cal N}=4$ SYM.  And, it is further highlighted
by the result that the ratios of the jet quenching parameters of 
different strongly coupled conformal theories are given by the 
ratios of the square roots of their entropy densities~\cite{Liu:2006he}.

\FIGURE[t]{
\includegraphics[width=14cm,angle=0]{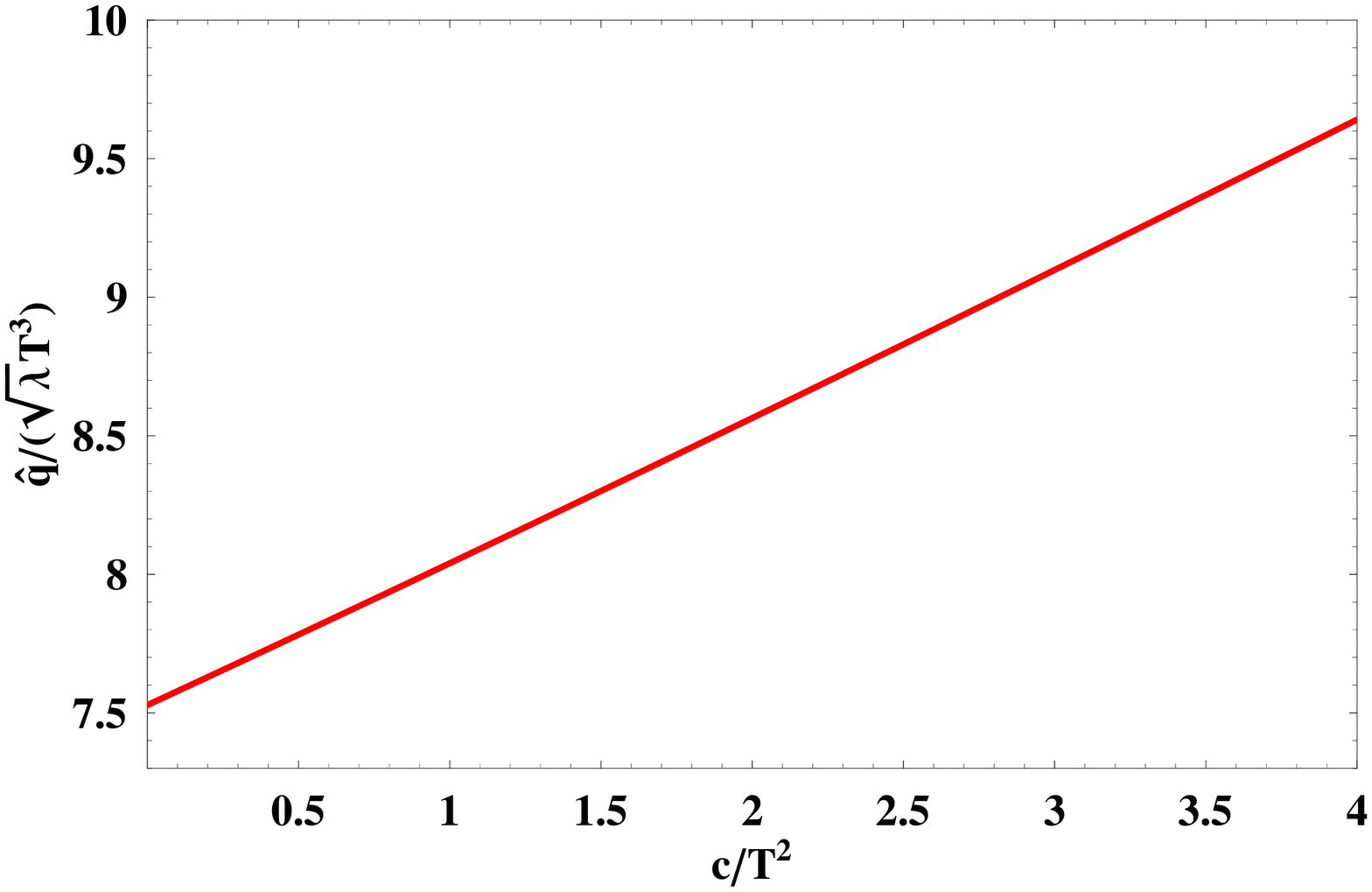}
\caption{The dependence of the jet quenching parameter
on the nonconformality in the metric (\protect{\ref{eq:fmetric}}). 
We plot $\hat{q}/(\sqrt{\lambda} T^{3})$ versus
$c/T^{2}$.}
\label{fig:hatq} }

In Fig.~\ref{fig:hatq}, 
we plot the dimensionless quantity
$\hat{q}/(\sqrt{\lambda} T^{3})$ against the
dimensionless quantity $c/T^2$, through which
nonconformality enters the calculation.  The dependence
on $c/T^2$ is almost linear over the range of $c/T^2$ that
is of  interest, and $\hat q$ increases only by about 28\% even
for the large value $c/T^2 = 4$.

\subsection{Robustness and Infrared Sensitivity}

Recall from Section 1 that the authors of the model (\ref{eq:fmetric})
find that $c=0.127$~GeV$^2$ best reproduces certain aspects of QCD
thermodynamics known from lattice calculations~\cite{Kajantie:2006hv}. And, recall
that we found
that the range $0.11$~GeV$^2 < c < 0.18$~GeV$^2$ yielded a
degree of nonconformality, parameterized by $(\varepsilon-3P)/\varepsilon$,
as in lattice QCD calculations.  By plotting $\hat q$ for values of $c/T^2$
up to 4, at $T=300$~MeV
we are allowing for values of $c$ at least twice as large as
is favored by QCD thermodynamics.   We see from Fig.~\ref{fig:hatq}
that even over this wide range of $c/T^2$, the jet quenching parameter
is at most increased by less than 30\%.  If we take $c=0.13$~GeV$^2$, 
the increase in $\hat q$ is $\sim 10\%$ at $T=300$~MeV
and $\sim 23$\% at $T=200$~MeV. We see first of all that the
${\cal N}=4$ SYM result is robust: upon
varying the degree of nonconformality
$c/T^2$ across a wide range, we
find only a small increase in $\hat q$.  Second of all, the fact that
$\hat q$ increases as we turn on $c/T^2$ is interesting.  
Among conformal theories,
if we reduce the number of degrees of freedom 
(with fermions weighted by a $7/8$ as in the entropy density) by a factor
of $47.5/120$, i.e. as if going from ${\cal N}=4$ SYM with $N_c=3$  to QCD, 
$\hat q$ is reduced by 
a factor of $\sqrt{47.5/120}\sim 0.63$~\cite{Liu:2006he}.
We now see that this decrease may be partially compensated by
an increase in $\hat q$ attributable to the nonconformality of QCD.
Our result that $\hat q$ increases with increasing nonconformality
has of course only been obtained in a particular toy model; further
investigation in other examples of 
nonconformal plasmas is called for.  One result that corroborates
the sign of the effect of nonconformality on $\hat q$ is the
determination that introducing nonzero $R$-charge chemical
potential(s) in ${\cal N}=4$ SYM, which introduces nonconformality, increases 
$\hat q$~\cite{Caceres:2006as}.  (See also Ref.~\cite{Vazquez-Poritz:2006ba}.)
There is one nonconformal strongly coupled plasma in
a (3+1)-dimensional gauge theory other than ${\cal N}=4$ SYM
for which $\hat q$ is known:  in the cascading
gauge theory of Refs.~\cite{Klebanov:2000hb,Buchel:2001gw},  
$\hat q/(\sqrt{\lambda}T^3)$ decreases with
decreasing temperature, which corresponds to simultaneously decreasing
number of degrees of freedom and increasing nonconformality~\cite{Buchel:2006bv}.
Further exploring (and separating) the effects of varying numbers of
degrees of freedom and varying degrees of nonconformality
on $\hat q$ requires calculating this quantity for other nonconformal
strongly coupled plasmas, for example that in ${\cal N}=2^*$ SYM~\cite{Buchel:2003ah}.
Certainly at present the indications are that all these effects only
modify $\hat q$ at the few tens of percent level, a robustness that
is supported by the present investigation of the effects of nonconformality
alone in a toy model.
If further study continues to support the idea that in going
from ${\cal N}=4$ SYM to QCD the jet quenching parameter 
decreases by a factor $\sqrt{47.5/120}\sim 0.63$ by virtue of the
decrease in degrees of freedom and increases by a few tens of percent
by virtue of the nonconformality of the QCD plasma at the temperatures
of about $(1.5-2) T_c$ explored at RHIC, then the 
observation~\cite{Liu:2006ug}
that $\hat q$ of ${\cal N}=4$ SYM theory
at $T=300$~MeV is in the same ballpark as
the range for the 
time-averaged $\bar{\hat q}$ extracted in comparison with RHIC 
data~\cite{Eskola:2004cr} 
will grow in importance.\footnote{Note also that in going from
RHIC to the LHC the dominant change in $\hat q$ will come from its
$T^3$ dependence. If we neglect any smaller changes
due to decreases in $\sqrt{\lambda}$ and the degree of
nonconformality, we predict that in going from RHIC to the
LHC the increase in $\hat q$ should be proportional to the
increase in multiplicity at mid-rapidity~\cite{Abreu:2007kv}.  That is,
we predict 
that the time-averaged $\bar{\hat q}$ extracted
in comparison with LHC data
should be greater than that extracted in comparison with  RHIC
data by the factor 
$(dN^{\rm LHC}/d\eta)/(dN^{\rm RHIC}/d\eta)$~\cite{Abreu:2007kv}.}

Our results confirm that $\hat q$ is an infrared sensitive quantity.  That
is, when we introduce $c/T^2 \neq 0$, modifying the AdS black hole metric at
scales of order $T$ but leaving it unmodified in the ultraviolet, we find that
$\hat q$ is affected by this modification.  This is consistent with the interpretation
of $\hat q$ as the parameter through which
the physics of the strongly coupled medium at scales
of order the temperature enters into the calculation of
radiative 
parton energy loss and jet quenching.  The infrared sensitivity of $\hat q$
comes about in its computation because in the gravity dual 
$\hat q$ is described by a string that extends all the way from the ultraviolet
regime  to the black hole horizon, probing the gauge theory at all scales
down to of order the temperature.\footnote{In addition to the
string worldsheet that determines $\hat q$, the light-like Wilson loop 
also bounds an extremal world sheet that explores the field theory
only on scales comparable to, and to the ultraviolet of, the Compton
wavelength of the test quark whose mass is taken to infinity in
defining the Wilson 
loop~\cite{Liu:2006ug,Argyres:2006vs,Liu:2006he,Argyres:2006yz,Argyres:2008eg}. When written
in terms of the parameter $z_0$, the action of this string is identical
for any metric that becomes the AdS metric (\ref{eq:AdSmetric}) 
asymptotically
in the ultraviolet~\cite{Argyres:2008eg}, meaning that it is infrared 
insensitive~\cite{Liu:2006he}.
As we discussed in Section 1,  in any theory 
that is described by a generic metric that becomes the AdS 
metric (\ref{eq:AdSmetric}) with parameter $z_0$ in the ultraviolet,  
any quantity that is specified in terms of $z_0$ rather than by
the temperature (which is determined by the metric near the
horizon and is in general not related to the ultraviolet
parameter $z_0$ in any simple way) is infrared insensitive. Thus,
the explicit calculations of Ref.~\cite{Argyres:2008eg} 
demonstrate quantitatively that this string solution only probes
physics at and beyond the ultraviolet cutoff and
is completely insensitive to physics of the strongly
coupled plasma at scales of order the temperature. The two
different string world sheets bounded by the light-like Wilson
loop can be thought of as different saddle points in its Minkowski-space
path integral representation.  
This is a Minkowski-space path integral, with an integrand
(proportional to $e^{iS}$) that is complex for real field configurations
(which have real $S$).  It is defined
by analytic continuation, with each of the integrals that
make up the path integral now a contour integral over the
complexified configuration space.  In this context,
there is no way to use the (imaginary) values
of the actions of the two saddle points to determine which dominates
the path integral. In the absence of information about
which saddle points lie on the infinite dimensional analogue
of the path of steepest descent, one must use physical considerations.
The calculations of Ref.~\cite{Argyres:2008eg} provide 
strong evidence confirming previous physical arguments:  
the infrared insensitive string world sheet 
does not contribute to the evaluation of the
Wilson loop, and hence $\hat q$~\cite{Liu:2006ug,Liu:2006he}.}


\section{Heavy quark drag and diffusion from AdS/CFT}

\subsection{Formulation}

The relativistic generalization of the Langevin equations for a
heavy quark moving through some thermal medium (see for example
Refs.~\cite{Moore:2004tg,Casalderrey-Solana:2006rq}) can be written
as
 \bea
 \label{lang1}
 {d p_L  \ov dt} & = & - \mu (p_L) p_L + \xi_L (t)\ , \\
 \label{lang2}
 {d p_T  \ov dt} & = &  \xi_T (t)\ ,
 \eea
where $p_L$ and $p_T$ are the longitudinal and transverse momentum
of the quark, respectively.  (We have simplified the notation by
dropping the spatial indices on transverse quantities.) 
Henceforth, we shall denote $p_L$ by $p$. $\xi_L$ and $\xi_T$ are random
fluctuating forces in the longitudinal and transverse directions,
which satisfy
 \bea
 \label{lang3}
 \vev{\xi_L (t) \xi_L (t')} & = & \kappa_L (p) \delta (t-t') \ ,\\
 \label{lang4}
 \vev{\xi_T (t) \xi_T (t')} & = & \kappa_T (p) \delta (t-t')\ .
 \eea
$\kappa_L (p)$ and two times $\kappa_T (p)$ describe how much
longitudinal and transverse momentum squared is transferred to the
quark per unit time. Note that  at zero velocity, $\kappa_L(0) =
\kappa_T (0)$ whereas for $p>0$ one expects that $\kappa_L(p)\neq
\kappa_T(p)$. Also, upon assuming that the momentum fluctuations
of the particle are in equilibrium with the thermal  medium, as
appropriate at zero velocity, a fluctuation-dissipation theorem
relates $\mu(0)$ to $\kappa_L(0)$ via the Einstein relation 
\be
\mu(0) = \frac{\kappa_L(0)}{2MT}\ , 
\label{Einstein} 
\ee 
where $M$
is the static mass of the quark. The relation (\ref{Einstein}) is
not expected to hold for $p>0$.

To compute the various quantities $\mu (p)$, $\kappa_T (p)$ and
$\kappa_L (p)$ in the metric
(\ref{eq:fmetric}), we use the following procedure developed
in Refs.~\cite{Casalderrey-Solana:2006rq,Gubser:2006nz,calsa}:

\bn

\item 
Find a classical solution to the Nambu-Goto action
\begin{equation}
S_{NG}=\frac{1}{2\pi\alpha'}\int d\tau d\sigma\sqrt{-\det
h_{\alpha\beta}} \label{eq:ng}
\end{equation}
which describes a trailing string moving with 
constant velocity~\cite{Herzog:2006gh,Gubser:2006bz} 
in
the metric~(\ref{eq:fmetric}).  Here, $h$ is the metric
induced on the string worldsheet.

\item 
The drag force is given by the momentum flux on the
worldsheet of the trailing string along the radial direction, i.e.~\cite{Herzog:2006gh,Gubser:2006bz}
 \be
 \frac{dp^i}{dt}= - {\delta S_{NG} \ov \delta \p_\sig x^i}
 \biggr|_{\rm trailing \;\; string} \ .
 \label{wsfl}
 \ee
 As we will see below, the right-hand side of (\ref{wsfl}) is a
 conserved quantity on the worldsheet and can be evaluated
 anywhere on the worldsheet.

\item 
Denote the retarded propagators for $\xi_L$ and $\xi_T$ as
$G_R^{(L)}$ and $G_R^{(T)}$ respectively. Then, 
the procedure for determining $\kappa_L$ and
$\kappa_T$ developed in Ref.~\cite{Gubser:2006nz,calsa} can be cast as
 \be \label{eowe}
 \kappa_{T, L} = -\lim_{\om \to 0} {2 T_{ws} \ov \om} G^{(T,L)}_R
 (\om)\ ,
 \ee
where $T_{ws}$ denotes the temperature on the worldsheet.
As we will see, the induced
metric on the trailing worldsheet has a horizon, meaning that the
worldsheet metric can be considered a $(1+1)$-dimensional 
black hole. $T_{ws}$
is the Hawking temperature for this worldsheet black hole. 
Note that at nonzero velocity $T_{ws}$ in general differs from the
temperature $T$ of the plasma itself. 
At
zero velocity, the worldsheet horizon coincides with that of the
spacetime, while at finite velocity the worldsheet horizon moves
closer to the boundary and the corresponding $T_{ws}$ decreases.
The reason that one should use the worldsheet temperature 
rather the
spacetime temperature in this computation is that the fluctuations
$\xi_T$ (and $\xi_L$) in the transverse (and longitudinal) directions
of the trajectory of the quark moving
through the gauge theory plasma
arise in the dual gravity description from
the fluctuations of the string worldsheet around the
trailing string solution~\cite{Gubser:2006nz,calsa}.  It is as if the force fluctuations
that the quark in the boundary gauge theory feels are due to
the fluctuations of the string worldsheet to which it is attached.

\item The retarded propagators $G_R^{(L,T)}$ can be found following
the general prescription given in~\cite{Son:2002sd}. One first
solves the linearized equation of motion for the worldsheet fluctuations with
the boundary conditions that they are infalling at the worldsheet
horizon and go to unity at the ultraviolet boundary. The retarded
propagator is then given by the action evaluated on this solution
(ignoring possible boundary terms at the horizon).

\en

\subsection{Finding the Trailing String and Calculating the Drag}

Consider a quark propagating in the
$x^{1}$ direction with velocity $v$. In this subsection 
we shall follow the analysis of Refs.~\cite{Herzog:2006gh,Gubser:2006bz} to
obtain the corresponding trailing string solution
and determine 
the drag force.

We parametrize the world sheet with $t$ and $r$ and use the ansatz 
\be 
x^{1}(t,r)=vt+\zeta(r)
\label{eq:DragAnsatz}
\ee
for a late-time steady-state solution.
The Nambu-Goto action (\ref{eq:ng}) is then
\begin{equation}
S_{NG}=\frac{1}{2\pi\alpha'}\int dtdr\LL
\label{eq:dfaction}
\end{equation}
with $\LL$ given by
\begin{equation}
\LL=e^{\frac{29}{20}c\frac{R^{4}}{r^{2}}}\sqrt{1+\frac{r^{4}-r_{0}^{4}}{R^{4}}\zeta'^{2}
-\frac{v^{2}r^4}{r^4- r_{0}^{4}  }}\ ,
\label{eq:dfactionansatz}
\end{equation}
where prime denotes differentiation with respect to $r$. The canonical momentum
\begin{equation}
\pi_{\zeta}\equiv\frac{\frac{r^{4}-r_{0}^{4}}{R^{4}}\zeta'}{\sqrt{1+\frac{r^{4}-r_{0}^{4}}{R^{4}}\zeta'^{2}
-\frac{v^{2}r^4}{r^4- r_{0}^{4}  }}}e^{\frac{29}{20}c\frac{R^{4}}{r^{2}}}
\label{eq:pixi}
\end{equation}
is conserved, meaning that
\begin{equation}
\zeta'=
\frac{R^{4} \pi_\zeta    }{r^{4}-r_{0}^{4}}
\sqrt{\frac{r^4- r_{0}^{4}  -v^{2}r^4 }{(r^4- r_{0}^{4})e^{\frac{29}{10}c\frac{R^{4}}{r^{2}}} 
- R^{4} \pi_{\zeta}^{2}
}} \ .
\label{eq:xieom}
\end{equation}
The integration constant $\pi_\zeta$ can be fixed by the following
argument:  
both the numerator and the denominator of the fraction
under the square root  in (\ref{eq:xieom}) are positive at
$r=\infty$ and negative at $r=r_{0}$;  since (\ref{eq:xieom})
is real, both must change sign at the same $r$; this is only the case if
\begin{equation}
\pi_{\zeta}=\frac{r_{0}^{2}v}{R^{2}\sqrt{1-v^{2}}}e^{\frac{29c\sqrt{1-v^{2}}R^{4}}{20r_{0}^{2}}}  ~.
\label{eq:pixifixed}
\end{equation}
The drag force (\ref{wsfl}) is then
\begin{align}
\frac{dp_{1}}{dt}
&=-\frac{\pi_{\xi}}{2\pi\alpha'} \nonumber \\
 & = -\frac{\sqrt{\lambda}\pi vT^{2}}
 {2\sqrt{1-v^{2}}}e^{\frac{29c\sqrt{1-v^{2}}}{20\pi^{2}T^{2}}} \ ,
\label{eq:dragforceworkedougt}
\end{align}
where we have used $R^{4}=\lambda\alpha'^{2}$ and $r_{0}=\pi
R^{2}T$ in the last step. 
The result (\ref{eq:dragforceworkedougt}) can also
be expressed in terms of the momentum $p_{1}$ and mass $M$ of the
external quark:
\begin{equation}
\frac{dp_{1}}{dt}=-\frac{\sqrt{\lambda}\pi T^{2}}{2}e^{\frac{29c\sqrt{1-v^{2}}}{20\pi^{2}T^{2}}}\frac{p_{1}}{M}  ~,
\label{eq:dragforcetwo}
\end{equation}
as obtained previously in Ref.~\cite{Nakano:2006js}.
We see that 
turning on $c$ increases the drag force, 
but the effect of the nonconformality becomes weaker at larger $v$. 
In fact, for $v\rightarrow 1$ the drag force is independent of $c$, meaning
that in this limit the drag force becomes an infrared insensitive observable.
The effects of the nonconformality are largest in the $v\rightarrow 0$ limit:
at low velocities, the drag force is increased  by a factor of 1.34 (1.80) 
for $c/T^2=2$ ($c/T^2=4$).  (As was also the case for $\hat q$,
the sign of the effect of nonconformality on the drag force 
is corroborated by the
determination that introducing nonzero $R$-charge chemical
potential(s) in ${\cal N}=4$ SYM, which introduces nonconformality, increases 
the drag force~\cite{Herzog:2006se}.)
Finally, notice that when $c$ is nonzero the
drag force is not proportional to the momentum. 
In other words, the drag coefficient 
$\mu(p_1) \equiv-\frac{1}{p_1}\frac{dp_{1}}{dt}$
now depends on the velocity and hence on the momentum $p_1$.


\subsection{Worldsheet Fluctuations}

The trailing string solution of Section 3.2
has $x^{2}=x^{3}=0$ and so after we change coordinates
from $r$ to $z=R^2/r$ it is specified by giving the dependence of
$x^1$ on $t$ and $z$.  Using (\ref{eq:DragAnsatz}) and (\ref{eq:xieom}),
this
can be written as
\begin{equation}
\frac{dx^{1}}{dt}=v 
\label{eq:unflutuatedt}
\end{equation}
and
\begin{equation}
\frac{dx^{1}}{d\bar{z}}=
-\frac{\bar{z}^{2}v}{1-\bar{z}^{4}}e^{\frac{29c(\sqrt{1-v^{2}}-\bar{z}^{2})}{20\pi^{2}T^{2}}}
\sqrt{\frac{1-v^{2}-\bar{z}^{4}}{1-v^{2}-\bar{z}^{4}\left(1-v^{2}+v^{2}
e^{\frac{29c(\sqrt{1-v^{2}}-\bar{z}^{2})}{10\pi^{2}T^{2}}}\right)}} ~,
\label{eq:unflutuatedz}
\end{equation}
where we have introduced $\bar{z}\equiv z/z_0$.

Following Ref.~\cite{calsa},
we now consider small fluctuations 
around the trailing string solution, which
we denote here by $x_0^1$, namely
\begin{equation}
x^{1}=x^{1}_{0}+\delta x^{1}(t,z)  ~,~~~~  
x^{2}=\delta x^{2}(t,z)  ~,~~~~  x^{3}=\delta x^{3}(t,z)  \ .
\label{eq:fluctuations}
\end{equation}
We expand 
the Nambu-Goto action (\ref{eq:ng}) to quadratic order in $\delta x^{i}$ to obtain
\begin{equation}
S_{NG}=S_{NG}^{0}+ \frac{R^{2}}{2\pi\alpha'} \int dtdz\,\le[
\GG_{L}^{\alpha\beta}\partial_{\alpha}\delta
x^{1}\partial_{\beta}\delta
x^{1}+\sum_{i=2,3}\GG_{T}^{\alpha\beta}\partial_{\alpha}\delta
x^{i}\partial_{\beta}\delta x^{i}\ri] \ ,\label{eq:quadratic}
\end{equation}
where $S_{NG}^{0}$ is the unperturbed action for the trailing string
solution.  The quantities $\GG_{T}^{\alpha\beta}$ and $\GG_{L}^{\alpha\beta}$ are
given by
 \be
 \label{Varde}
 \GG_T^{\al \beta} = f_T \sqrt{-h} h^{\al \beta}, \qquad
  \GG_L^{\al \beta} = f_L \sqrt{-h} h^{\al \beta}\ ,
 \ee
where $h_{\al \beta}$ is the induced worldsheet metric 
whose components we can evaluate using 
(\ref{eq:unflutuatedt}) and (\ref{eq:unflutuatedz}), obtaining
 \begin{align}
h_{tt} =-\frac{R^2 A\sqrt{1-v^{2}}}{\hat{z}^{2}}e^{\frac{29c\sqrt{1 - v^2}\hat{z}^{2}}{20\pi^{2}T^{2}}} ~,
\label{hwstt}
\end{align}
\begin{equation}
h_{tz} =h_{zt} =-\frac{R^2 v^{2}e^{\frac{29c\sqrt{1-v^{2}}}{20\pi^{2}T^{2}}}}{1-(1-v^2)\hat{z}^{4}}\!\sqrt{\frac{A}{B}} ~,
\label{hwstz}
\end{equation}
and
\begin{equation}
h_{zz}=\frac{-e^{\frac{29c\sqrt{1-v^2}(1-\hat{z}^{2})}{10\pi^{2}T^{2}}}v^{4}\hat{z}^{4}
+\left[1-(1-v^2)\hat{z}^{4}\right]^{2}  }{\hat{z}^2\sqrt{1-v^2}
\left[1-(1-v^2)\hat{z}^{4}\right]^{2} B}
R^2e^{\frac{29c\sqrt{1-v^{2}}\hat{z}^{2}}{20\pi^{2}T^{2}}} \ ,
\label{eq:hwszz}
\end{equation}
where we have introduced
\be
\hat z \equiv \sqrt{\ga} \bar{z}=\sqrt{\gamma}z/z_0=\sqrt{\gamma}z\pi T\ ,
\ee
with $\gamma\equiv  1 / \sqrt{1-v^{2}}$ and defined
\begin{align}
A\equiv 1-\hat{z}^{4} ~,
\label{eq:shorthandA}
\end{align}
\begin{align}
B\equiv 
1-\hat{z}^{4}\left[1-\left(1-e^{\frac{29c\sqrt{1-v^2}(1-\hat{z}^{2})}{10\pi^{2}T^{2}}}\right)v^{2}\right]
\ ,
\label{eq:shorthandB}
\end{align}
and where the prefactors in (\ref{Varde}) are
given by
 \bea
f_T & \equiv& \frac{e^{\frac{29\,c\,{\sqrt{1 -
v^2}}\,{\hat{z}}^2}{20\,{\pi }^2\,T^2}}}
  {2\,{\sqrt{1 - v^2}}\,{\hat{z}}^2\,
     } \\
f_L & \equiv & \frac{B}{(1- v^2)\,A} f_T \ .
 \eea

We now make a change of worldsheet coordinates that
diagonalizes
the worldsheet metric $h_{\al \beta}$.   This will simplify the calculation
since, as is clear from (\ref{Varde}), diagonalizing $h_{\al \beta}$
will also automatically diagonalize $\GG_T$ and $\GG_L$. 
For convenience, we first change coordinates from $z$ to $\hat z$.
Then, we define a new coordinate
  \be
  \hat t =  t + g(\hat z)
  \ee
where $g(\hat z)$ 
satisfies 
\begin{align}
\frac{\partial g}{\partial \hat z} =
\frac{v^2\,{\hat{z}}^2\,e^{\frac{29\,c\,{\sqrt{1 - v^2}}\,\left( 1 - {\hat{z}}^2 \right) }
       {20\,{\pi }^2\,T^2}}}
  {\left( 1 - v^2 \right)^{\frac{1}{4}}\,
    \left[ 1 - \left( 1 - v^2 \right) \,{\hat{z}}^4 \right]\! \sqrt{AB} } \ ,
\label{eq:ct}
\end{align}
which ensures that 
$h_{\hat t \hat z}$ vanishes.
In the new $(\hat z,\hat t)$ worldsheet coordinate system, the induced worldsheet 
metric $h_{\hat{\alpha}\hat{\beta}}$ becomes 
\begin{align}
h_{\hat{t}\hat{t}}&=-\frac{R^2 A}{\ga \hat{z}^{2}}e^{\frac{29c\sqrt{1-v^{2}}\hat{z}^{2}}{20\pi^{2}T^{2}}} ~, 
\label{hhatthatt}\\
h_{\hat{z}\hat{z}}&=\frac{R^2}{\hat{z}^{2}\,B}e^{\frac{29c\sqrt{1-v^{2}}\hat{z}^{2}}{20\pi^{2}T^{2}}} ~.
\label{hhatzhatz}
\end{align}
We now see that $h_{\hat{t}\hat{t}}$ vanishes and
$h_{\hat{z}\hat{z}}$ diverges at $\hat z=1$. This demonstrates
that the induced metric on the (1+1)-dimensional
worldsheet has an event horizon at $\hat{z}=1$, corresponding
to $z=z_0/\sqrt{\gamma}=1/(\pi T \sqrt{\gamma})$ 
and $r=r_0 \sqrt{\gamma} = R^2 \pi T \sqrt{\gamma}$. 
Note that the worldsheet horizon moves toward the ultraviolet as $v\rightarrow 1$ and 
$\gamma\rightarrow\infty$.
The $\hat{z}<1$ region of the worldsheet is outside, and
to the ultraviolet of, the worldsheet horizon.  The $\hat z > 1$
region is inside, and to the infrared. 
Classically, no signal from the interior of the worldsheet
horizon can propagate along the worldsheet 
to the exterior. The Hawking temperature $T_{ws}$ of the
worldsheet black hole is obtained as follows. First, we note that via 
a change of coordinates, the worldsheet metric outside
the horizon, in the vicinity
of the horizon,  takes the form
$ds^2 = - b^2 \rho^2 d\hat{t}^2 + d\rho^2$ for some constant $b$,
where the horizon is at $\rho =0$.  Then, it is a standard argument
that in order to avoid having a conical singularity at $\rho=0$ in 
the Euclidean version of this metric, $b \hat{t}$ must be a periodic
with period $2\pi$.  We then identify
the  period of $\hat{t}$, namely $2\pi/b$, as $1/T_{ws}$.
This argument yields
\begin{equation}
T_{ws}=\frac{T}{\sqrt{\gamma}} \,
\sqrt{1-\frac{29cv^{2}\sqrt{1-v^{2}}}{20\pi^{2}T^{2}}}\ . 
\label{eq:wsbhtunits} 
\end{equation}
The diffusion in momentum space of the moving heavy quark, governed
by the diffusion constants $\kappa_T$ and $\kappa_L$, 
is described in the
dual gravity theory by
the fluctuations of the worldsheet 
outside the worldsheet horizon due to the worldsheet Hawking
radiation with temperature $T_{ws}$.  With
$T_{ws}$ in hand, we turn now to the calculation of 
the diffusion constants (\ref{eowe}).

\subsection{Calculation of $\kappa_{T}$}

We now 
calculate the two point function for the transverse
fluctuations, starting from the quantity $\GG_T$ defined in
(\ref{eq:quadratic}) and given explicitly
in Eqs.~(\ref{Varde})-(\ref{eq:shorthandB}) and (\ref{hhatthatt}) and
(\ref{hhatzhatz}).  It turns out that it is
convenient to define $u\equiv\hat{z}^{2}$ as the radial coordinate in
the calculations of $\kappa_{T}$ and $\kappa_{L}$. 
We write the transverse fluctuations
part of the action as 
\begin{equation}
S_{NG}^{T}=\frac{R^{2}}{2\pi\alpha'}\int dt\,du\,[\,g_1 (\p_t \delta
y)^{2}+ g_2(\p_u \delta y)^{2}\,]\ , \label{eq:ngatf}
\end{equation}
where $\delta y$ here can be either $\delta x_2$ or $\delta x_3$
and where
\begin{align}
g_1 &\equiv\frac{\GG_{T}^{\hat{t}\hat{t}}}{2\sqrt{u}}=
-\frac{1}{4\,u^{\frac{3}{2}}\,\left( 1 - u^2 \right) \,(1 -
v^2)^{\frac{3}{4}}\! \sqrt{AB}} e^{\frac{29\,c\,u\,{\sqrt{1 - v^2}}}{20\,{\pi }^2\,T^2}}
 ~,
\label{eq:gtttran}
\end{align}
\begin{align}
g_2 &\equiv2\sqrt{u}\GG_{T}^{\hat{z}\hat{z}}= \frac{\sqrt{AB}}{{\sqrt{u}}\,(1 - v^2)^{\frac{1}{4}}} e^{\frac{29\,c\,u\,{\sqrt{1 -
v^2}}}{20\,{\pi }^2\,T^2}} ~,
\label{eq:guutran}
\end{align}
and where it is understood that we have
rewritten A from (\ref{eq:shorthandA}) and B from (\ref{eq:shorthandB}) 
in terms of $u$, obtaining
\begin{equation}
A=1 - u^2 ~, 
\end{equation}
\begin{equation}
B=1 - \left[ 1 - \left( 1 - e^
         {\frac{29\,c\,( 1 - u ) \,{\sqrt{1 - v^2}}}{10\,{\pi }^2\,T^2}}
       \right) \,v^2 \right] \, u^2  ~.
\end{equation}
The equation of motion for the transverse fluctuations $\delta y$ is given by
\begin{equation}
\partial_{u}^{2}\delta y+\frac{\partial_{u}g_2}{g_2}\partial_{u}\delta y
+\frac{g_1}{g_2}\partial_{t}^{2}\delta y=0 ~. \label{eq:eleqtrans}
\end{equation}
After a Fourier transformation
\begin{equation}
\delta y(t,u)=\int_{-\infty}^{\infty}e^{-i\omega t}Y_{\omega}(u)\frac{d\omega}{2\pi} ~,
\label{eq:ft}
\end{equation}
the equation of motion (\ref{eq:eleqtrans}) becomes
\begin{equation}
\partial_{u}^{2}Y_{\omega}+
\frac{\partial_{u}g_2}{g_2}\partial_{u}Y_{\omega}-\frac{\omega^{2}g_1}{g_2}Y_{\omega}=0
~. \label{eq:eleqtransfourier}
\end{equation}
To examine the behavior of the solution near the worldsheet
horizon $u=1$, we expand the coefficients in
(\ref{eq:eleqtransfourier}) near $u=1$ and obtain
\begin{equation}
\partial_{u}^{2}Y_{\omega}+\frac{1}{u-1}\partial_{u}Y_{\omega}+\frac{\gamma\,{\omega }^2}
  {8\,{( u-1 ) }^2\,
    \left( 2 - \frac{29\,c\,v^2\,{\sqrt{1 - v^2}}}
       {10\,{\pi }^2\,T^2} \right) }Y_{\omega}=0 ~,
\label{eq:indexeq}
\end{equation}
whose solution is
\begin{equation}
Y_{\omega}=(1-u)^{\pm\frac{i\,\sqrt{\gamma}\,\omega}{2\sqrt{4-\frac{29cv^{2}\sqrt{1-v^{2}}}{5\pi^{2}T^{2}}}}} ~.
\label{eq:solutionnearone}
\end{equation}
For the solution with the ``plus'' sign, the phase increases as one goes to smaller value of $u$, i.e. 
``outward'' from the worldsheet horizon, toward the ultraviolet,
meaning that this corresponds to an
outgoing solution, which is to be discarded in our case. We only need the infalling solution. Therefore, we can write
our solution $Y_{\omega}$ as
\begin{equation}
Y_{\omega}=(1-u^{2})^{-\frac{i\,\sqrt{\gamma}\,\omega}{2\sqrt{4-\frac{29cv^{2}\sqrt{1-v^{2}}}{5\pi^{2}T^{2}}}}}F(\omega,u) ~,
\label{eq:factorout}
\end{equation}
where $F(\omega,u)$ is regular at the horizon. We now substitute
(\ref{eq:factorout}) into (\ref{eq:eleqtransfourier}) and obtain an
ordinary differential equation for $F$ that takes the form
\begin{equation}
X\,F+V\,\partial_{u}F+\partial_{u}^{2}F=0 ~,
\label{eq:odeforF}
\end{equation}
where $X$ and $V$ are functions of $u$ and $\omega$ (that
depend on $v$ and $T$) whose leading behavior at small $\omega$
is given explicitly in Appendix~\ref{ap:A}.

In order to determine $\kappa_T$, we only need
to find the solution $F$ to (\ref{eq:odeforF}) to first order in
$\omega$. We show in Appendix 
\ref{ap:A} that to zeroth order in $\omega$ the only solutions
that are regular at the horizon at $u=1$ are $F=$constant.
We normalize $Y_\om$ so that 
$Y_\om \rightarrow 1$ at $u\rightarrow 0$, and this determines
that we choose $F=1$ to zeroth order in $\omega$.
To first order in $\omega$
the solution then takes the form
\begin{equation}
F=1+\omega Z + O(\om^2)\ ,
\label{eq:fomegaz}
\end{equation}
and in Appendix \ref{ap:A} we show that 
the function $Z$ has the properties
that it goes to a constant at the
horizon $u=1$ and that
 \be \label{z3}
   Z \to
\frac{i}{3}\,\sqrt{\gamma}\,e^{\frac{29c\sqrt{1-v^{2}}}{20\pi^{2}T^{2}}}
 u^{3 \ov 2} + \cdots
 \ee
as $u \to 0$.
Upon normalizing $Y_\omega$ at 
$u\rightarrow 0$ as we have done, the retarded propagator
that appears in (\ref{eowe}) is given by~\cite{Son:2002sd}
 \be
G_{R}^T(\omega) \,=\,  -\frac{R^{2}(\pi
T)^{2}}{\pi\alpha'}g_{2}\,Y_{-\omega}(u)\partial_{u}Y_{\omega}(
u)\Bigr|_{u\rightarrow 0} 
 \,=\, - \sqrt{\lam} \pi  T^{2} g_{2}\, \om \p_u Z (u) \Bigr|_{u\rightarrow
 0}+ O(\om^2)\ ,\label{eq:GRintermsofZ}
\ee
where $g_2$ is given in (\ref{eq:guutran}).
Using (\ref{z3}) and the fact that 
$g_{2}=\frac{1}{(1-v^{2})^{\frac{1}{4}}\sqrt{u}}+\OO(\sqrt{u})$ in
the $u\rightarrow 0$ limit, 
we find that with the propagator (\ref{eq:GRintermsofZ}) and
the world sheet temperature (\ref{eq:wsbhtunits}) the
transverse momentum diffusion constant (\ref{eowe}) is given by
\begin{equation}
\kappa_{T}=\alpha_T\sqrt{\gamma}\sqrt{\lambda}\pi T^{3}
\label{eq:ktanswer}
\end{equation}
where 
\be
\alpha_T = 
e^{\frac{29c\sqrt{1-v^{2}}}{20\pi^{2}T^{2}}}\sqrt{1-\frac{29cv^{2}\sqrt{1-v^{2}}}{20\pi^{2}T^{2}}} ~.
\label{eq:alpha_T}
\ee
When $c=0$, $\alpha_T=1$ and our result reduces to the
known result for
${\cal N}=4$ SYM, derived in  Refs.~\cite{Gubser:2006nz,calsa}.
From our result, we see that turning on $c/T^2$ increases $\kappa_T$ by
a factor $\alpha_T$, which we have plotted in Fig.~\ref{fig:alpha_T} as 
a function of $v$ for several values of $c/T^2$.  Comparing $\alpha_T$ 
at $v=0$ from (\ref{eq:alpha_T}) to our result (\ref{eq:dragforcetwo}) for the drag coefficient
evaluated at $v=0$, we see that at $v=0$ the Einstein relation (\ref{Einstein}) is valid 
with $c\neq 0$. This
can be seen as a consistency check on the model.

\FIGURE[t]{
\includegraphics[width=14cm,angle=0]{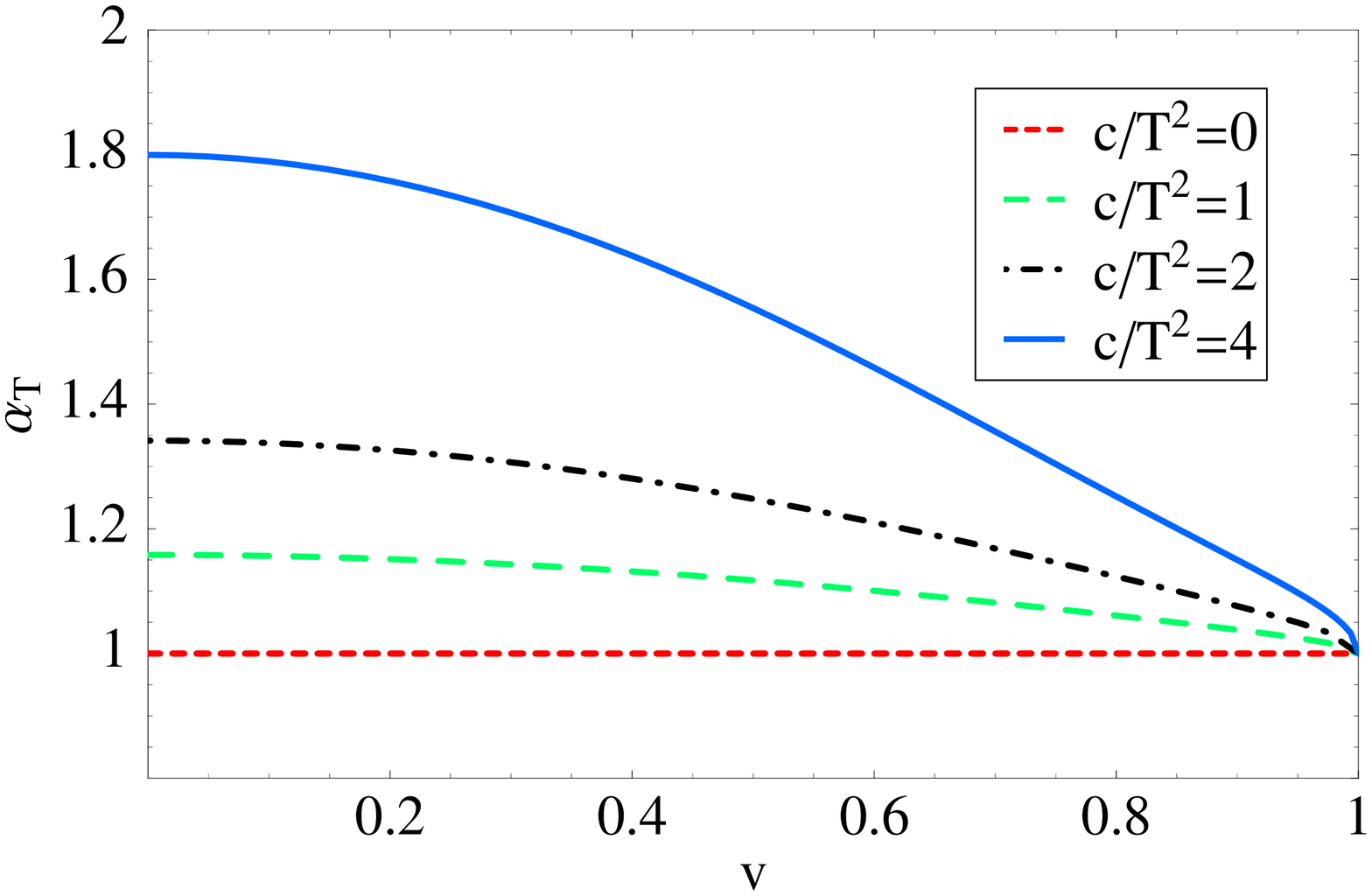}
\caption{The modification of $\kappa_T$
due to nonconformality is given by $\alpha_T$ in (\protect{\ref{eq:alpha_T}}),
which we plot here  versus $v$ at four values of $c/T^2$.
}
\label{fig:alpha_T}
}

We see from Fig.~\ref{fig:alpha_T} that
the effect of $c/T^2=4$ ($c/T^2=2$) on $\kappa_T$ at low velocity 
is significant, increasing it by a factor of 1.80  (1.34).
This suggests that the nonconformality
in QCD could increase the diffusion
constant $\kappa_T$, which has been related to charm 
quark energy loss and azimuthal anisotropy in 
Refs.~\cite{Moore:2004tg,Adare:2006nq}, by a 
significant factor relative to estimates based upon 
the ${\cal N}=4$ SYM result.
By comparing Fig.~\ref{fig:alpha_T} to Fig.~\ref{fig:hatq}, we also 
see that at low 
velocities $\kappa_T$ 
is less robust with respect to the introduction of nonconformality
than $\hat q$:  the effects of $c/T^2$ on $\kappa_T$ 
at low velocity are more than a factor of two   larger than its
effects on $\hat q$.
However, we also see that $\alpha_T\rightarrow 1$ 
for $v\rightarrow 1$: at high velocities,
$\kappa_T$ does not 
sense the nonconformality at all. 
This infrared insensitivity in the high velocity 
regime can be understood immediately once we recall
that the worldsheet horizon is at
$z=z_0 (1-v^{2})^{1/4}$, and $\kappa_{T}$ only depends on the 
portion of the string worldsheet that is outside
the horizon, namely between the ultraviolet boundary at $z=0$ and
$z=z_0 (1-v^{2})^{1/4}$.
At high velocity, $z_0 (1-v^{2})^{1/4}$ itself 
moves closer and closer
toward the boundary, i.e. farther and farther into the ultraviolet, meaning 
that $\kappa_T$ only probes ultraviolet physics where
$c$ is not important.

\subsection{Calculation of $\kappa_{L}$}

The calculation of $\kappa_{L}$ is analogous to that of
$\kappa_{T}$. The relevant action for longitudinal fluctuation takes
the same form as (\ref{eq:ngatf}), with $g_1$ and $g_2$ replaced
by
\begin{equation}
g_1=-\frac{1}{4\,u^{\frac{3}{2}}\,\left( 1 - u^2 \right) \,{\left(
1 - v^2 \right) }^{\frac{7}{4}}}e^{\frac{29\,c\,u\,{\sqrt{1 -
v^2}}}{20\,{\pi }^2\,T^2}}\,
  {\sqrt{\frac{B}{A}}} ~,
\label{eq:gttlong}
\end{equation}
\begin{equation}
g_2=\frac{1 - u^2}{{\sqrt{u}}\,{\left( 1 - v^2 \right)
}^{\frac{5}{4}}}e^{\frac{29\,c\,u\,{\sqrt{1 - v^2}}}{20\,{\pi
}^2\,T^2}}\,
  {\left( \frac{B}{A} \right) }^{\frac{3}{2}} ~.
\label{eq:guulong}
\end{equation}
The solution again takes the form (\ref{eq:factorout}),
with $F$ satisfying (\ref{eq:odeforF}), but with different
expressions for the functions
$X$ and $V$, given in Appendix A.2.
The expansion (\ref{eq:fomegaz}) still holds,  but now $Z(u)$
can only be obtained numerically. Again, as
described in Appendix \ref{ap:A}, 
the solution for $Z$ is determined uniquely by requring
that $Z$ be regular at $u=1$ and that $Z\rightarrow 0$,
so that $Y_\omega \rightarrow 1$,
as $u\rightarrow 0$.
We find that $\kappa_{L}$ is given by
\begin{equation}
\kappa_{L}=2\gamma^{-\frac{1}{2}}\sqrt{\lambda}\pi
T^{3}\sqrt{1-\frac{29cv^{2}\sqrt{1-v^{2}}}{20\pi^{2}T^{2}}}\,g_{2}\,
\left(\frac{\p_u Z}{i}\right)\Biggr|_{u\rightarrow 0} ~. \label{eq:klw}
\end{equation}
Near $u=0$, $g_{2}$ can be expanded 
as $g_{2}=\frac{1}{(1-v^{2})^{\frac{5}{4}}\sqrt{u}}+\OO(\sqrt{u})$, which reduces (\ref{eq:klw}) to
\be
\kappa_{L}
=\alpha_L\gamma^{\frac{5}{2}}\sqrt{\lambda}\pi T^{3} ~,
\label{eq:klwmore}
\ee
where
\be
\alpha_L=\frac{2}{\sqrt{\gamma}}\sqrt{1-\frac{29cv^{2}\sqrt{1-v^{2}}}{20\pi^{2}T^{2}}}
\left(\frac{\p_u Z}{i\,\sqrt{u}}\right)\Biggr|_{u\rightarrow 0}\label{eq:alpha_L}
\ee
depends on $v$ and $c/T^{2}$. 
We have checked analytically that $\alpha_L=1$ for $c=0$, 
meaning that our result reduces to that for ${\cal N}=4$ SYM
from Refs.~\cite{Gubser:2006nz,calsa} when the nonconformality is turned off.
That is, $\alpha_L$ is the factor by which $\kappa_L$ is
modified when we introduce nonconformality via nonzero $c/T^2$.
At nonzero
$c/T^2$, we compute $Z(u)$ and hence $\alpha_L$
numerically. 
In Fig.~\ref{fig:alphaV} we plot $\alpha_L$ versus $v$
at four values of $c/T^2$.  
The factor $\alpha_L$ is comparable
to but somewhat less than its counterpart $\alpha_T$ for the transverse
momentum diffusion constant $\kappa_T$, plotted in Fig.~\ref{fig:alpha_T}.
As $v\rightarrow 0$, $\alpha_L$ and $\alpha_T$ 
become equal because there is no difference between (diffusion in)
longitudinal and transverse momentum when $v=0$.
The effects of $c/T^2$ on $\kappa_L$ at small velocity are more than twice as large as
its effects on $\hat q$, but $\kappa_L$ becomes completely unaffected
by $c/T^2$, namely infrared insensitive, as $v\rightarrow 1$.  
We also see that there is a range of velocities near 1 for which $\alpha_L<1$.

\FIGURE[t]{
\includegraphics[width=14cm,angle=0]{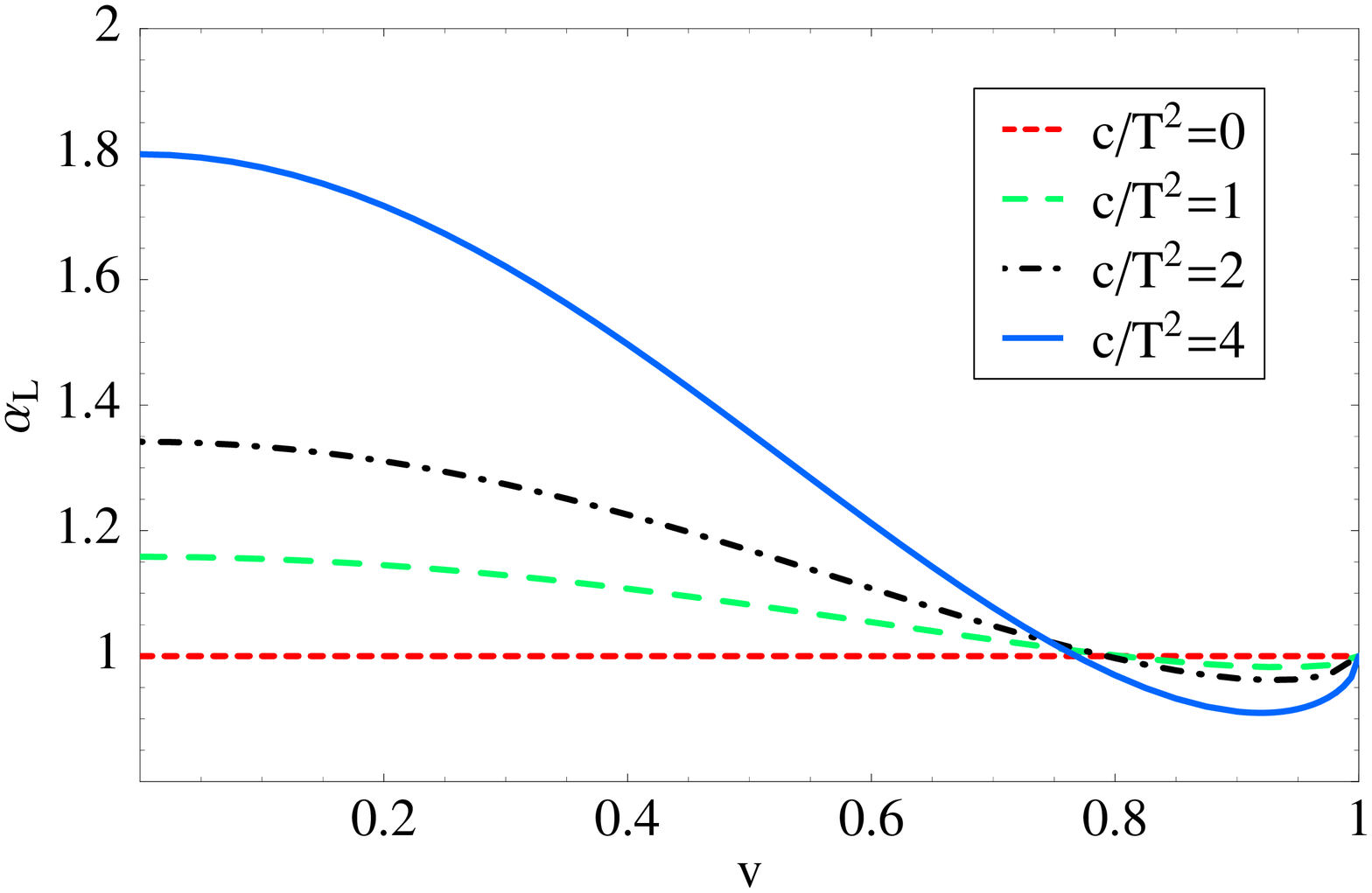}
\caption{The modification of $\kappa_L$
due to nonconformality is given by $\alpha_L$ in (\protect{\ref{eq:alpha_L}}),
which we plot here  versus $v$ at four values of $c/T^2$.
}
\label{fig:alphaV}
}

\subsection{Robustness and Infrared Sensitivity}

The effects of the nonconformality we have introduced on
all three of the quantities that we have computed that describe
the drag and diffusion of a heavy quark moving through the
plasma are comparable at low velocities.  For $c/T^2 = 1,2,4$, the
nonconformality serves to increase all three quantities
that we have computed, by factors of 1.16, 1.34, 1.80 at $v=0$.
So, particularly at lower temperatures where $c/T^2$
is larger, the tendency for the drag and diffusion coefficients
to increase with nonconformality should be included in
making estimates of these quantities for the QCD plasma.
We also showed that 
the energy loss on a heavy quark moving
through the plasma with $v\neq 0$ is not described
by a velocity-independent drag coefficient.  If
we define the drag coefficent $-\frac{1}{p}\frac{dp}{dt}$ 
then this quantity depends significantly 
on the velocity of the quark.

All three quantities that we have computed in this section
become completely infrared {\it in}sensitive for $v\rightarrow 1$.
The ${\cal N}=4$ SYM results for $\mu$, $\kappa_T$ and $\kappa_L$ 
are conventionally
quoted in terms of the temperature $T$, but their infrared insensitivity
for $v\rightarrow 1$ demonstrates that in this regime they should
really be quoted in terms of $z_0$, with $z_0$ understood as a
parameter that specifies the asymptotic ($z\rightarrow 0$) behavior
of the metric.   In a generic metric that is given asymptotically 
by the AdS black hole metric (\ref{eq:AdSmetric}),
$z_0$ is not related in any simple way to the temperature $T$, which
is determined by the metric in the vicinity of the horizon.
And, in the $v\rightarrow 1$ regime,
$\mu$, $\kappa_T$ and $\kappa_L$ are determined
by $z_0$ not by $T$.  The reason for the infrared insensitivity of all three
quantities is the same. The drag and diffusion
of the quark is described by that segment of the attached string
worldsheet that is outside the worldsheet event horizon at 
$z=z_0 (1-v^2)^{\frac{1}{4}}$.
As $v$ approaches 1, this
worldsheet event horizon moves to smaller and smaller $z$, meaning that
the segment of the worldsheet that is outside its event horizon,
namely at $z<z_0(1-v^2)^{\frac{1}{4}}$,
explores the metric only at smaller and smaller $z$, meaning that
it describes the physics of the
more and more ultraviolet sector of the gauge theory.
Because the metric that we are using is given asymptotically
at small $z$ by the AdS black hole metric (\ref{eq:AdSmetric}),
independent of $c$, 
the drag and diffusion of the quark become completely insensitive
to $c/T^2$ for $v\rightarrow 1$.  That is, they become infrared insensitive,
probing the gauge theory only at more and more ultraviolet scales.
We saw in Section 2 that, in contrast, the jet quenching parameter $\hat q$,
which is defined at $v=1$, is infrared sensitive.


\section{Quark-Antiquark Potential and Screening Length}

One of the early, classic, computations done using the
AdS/CFT correspondence was the computation of the potential
between a static quark and antiquark separated by a distance $L$
in ${\cal N}=4$ SYM theory,
first at zero temperature where the potential is Coulomb-like,
proportional to $\sqrt{\lambda}/L$~\cite{Rey:1998ik}, 
and then at nonzero temperature~\cite{Rey:1998bq}, 
where, to order $\sqrt{\lambda}$ in the
strong coupling expansion, the screened
potential is Coulomb-like for
$L\ll L_c(T)$ and flat for $L\gg L_c(T)$ (up to 
order $\lambda^0$ contributions that fall exponentially with $L$~\cite{Bak:2007fk}).
The screening length
turns out to be $L_c= 0.24/T$. When $L<L_c(T)$, the potential
is determined to order $\sqrt{\lambda}$ at strong coupling
by the area of a string worldsheet bounded by
the worldlines of the quark and antiquark, with the worldsheet
``hanging'' into the AdS black hole spacetime (\ref{eq:AdSmetric}),
``suspended''
from the test quark and antiquark that are located at the
ultraviolet boundary at $z=0$.

In Refs.~\cite{Liu:2006nn,Liu:2006he}, the analysis 
of screening was extended to the case of 
a quark-antiquark pair moving through the plasma
with velocity $v$.  In that context, it proved convenient
to define a slightly different screening length $L_s$,
which is the $L$ beyond which no connected 
extremal string
world sheet hanging between the quark and antiquark 
can be found.  At $v=0$, $L_s=0.28/T$~\cite{Rey:1998bq}.
At nonzero $v$, up to small corrections 
that have been computed~\cite{Liu:2006nn,Liu:2006he},
\be
 L^{\rm meson}_{s} (v,T) \simeq L_s (0,T) (1-v^2)^{1/4} 
 \propto \frac{1}{T}(1-v^2)^{1/4}\ .
  \label{lmax}
 \ee
This result, also obtained in~\cite{Peeters:2006iu,Chernicoff:2006hi} and further
explored in~\cite{Caceres:2006ta,Avramis:2006em,Natsuume:2007vc}, 
has proved robust
in the sense that it applies in various strongly
coupled plasmas other than 
${\cal N}=4$ SYM~\cite{Caceres:2006ta,Avramis:2006em,Natsuume:2007vc},
including some which are made nonconformal via the introduction
of R-charge chemical potentials.
The robustness
of the result (\ref{lmax}) has been tested in a second sense
by analyzing the potential and screening length defined by
a configuration consisting of $N_c$ external quarks arranged in
a circle of radius $L$, a ``baryon'', moving
through the strongly coupled plasma~\cite{Athanasiou:2008pz}.  In order to
obtain a baryon-like configuration, the $N_c$ strings hanging
down into the AdS black hole spacetime must end on a D5-brane~\cite{Witten:1998xy}.
Even with this qualitatively new ingredient, a screening length
once again emerges naturally, and obeys (\ref{lmax}) for ``baryons''
moving through the plasma~\cite{Athanasiou:2008pz}.
The velocity dependence of the screening
length (\ref{lmax}) suggests that in
a theory containing dynamical heavy quarks and meson
bound states (which $\NN=4$ SYM does not) 
the dissociation 
temperature $T_{\rm diss}(v)$, 
defined as the temperature above which mesons with
a given velocity do not exist,
should scale with velocity as~\cite{Liu:2006nn}
 \be \label{rro}
 T_{\rm diss} (v) \simeq  T_{\rm diss} (v=0) (1-v^2)^{1/4} \ ,
   \ee
since $T_{\rm diss}(v)$ should be the temperature at which
the screening length $L^{\rm meson}_s(v)$ is comparable to the size of
the meson bound state.  
The scaling (\ref{rro}) 
indicates that slower mesons can exist up to
higher temperatures than faster ones, a result which
has observable consequences for charmonium (bottomonium) production
as a function of transverse momentum in heavy ion collisions
at RHIC (LHC)~\cite{Liu:2006nn,Liu:2006he}.
This result has proved robust in a third sense, in that (\ref{rro})  
has also been obtained 
by direct analysis of the dispersion relations 
of actual mesons in 
the plasma~\cite{Mateos:2007vn,Ejaz:2007hg}, 
introduced by adding heavy quarks described
in the gravity dual by a D7-brane whose fluctuations are 
the mesons~\cite{Karch:2002sh,Mateos:2007vn,Ejaz:2007hg}.
These mesons have a limiting velocity whose temperature dependence 
is equivalent to (\ref{rro}), up to few percent corrections that have been 
computed~\cite{Ejaz:2007hg}.  

In this section, we shall return to the velocity-dependent
screening length defined by a quark-antiquark pair moving through
the plasma and test the robustness of (\ref{lmax}) 
in yet one more way by repeating
the calculation of $L_s(v,T)$ from Refs.~\cite{Liu:2006nn,Liu:2006he} in the
metric (\ref{eq:fmetric}) that incorporates
the nonconformal deformation whose consequences we are exploring
throughout the present paper.

\subsection{Calculating the Potential}

Consider an external 
quark-antiquark dipole moving with velocity $v=\tanh\eta$,
where $\eta$ is the rapidity of the dipole,  along the
$-x_{3}$ direction. We choose the quark and antiquark to
be separated by a distance $L$, oriented in the 
$x_{1}$ direction.
It proves convenient to boost into a frame in which the dipole
is at rest in a moving medium --- it feels a ``hot wind'' --- via
a Lorentz transformation
$(t,x_{3})\rightarrow(t', x_{3}')$:
\begin{eqnarray}
dt=&dt'\cosh\eta-dx_{3}'\sinh\eta \\
dx_{3}=&-dt'\sinh\eta+dx_{3}'\cosh\eta\ .
\end{eqnarray}
In the dipole rest frame, the spacetime metric describing
the nonconformal hot wind is obtained by applying the Lorentz
transformation to the metric
(\ref{eq:fmetric}), obtaining
\begin{align}
ds^{2}=&\frac{R^{2}}{z^2} e^{\frac{29cz^{2}}{20}}
 \Biggl\{\Biggl[ \sinh^{2}\eta
-\Biggl(1-\frac{z^{4}}{z_{0}^4}\Biggr)\cosh^{2}\eta\Biggr] dt'^{2}+
\Biggl[\cosh^{2}\eta
-\Biggl(1-\frac{z^{4}}{z_{0}^4}\Biggr)\sinh^{2}\eta\Biggr]dx_{3}'^{2}
\nonumber\\
&\qquad\qquad -2\frac{z^{4}}{z_{0}^4}\sinh\eta\cosh\eta
dt'dx_{3}'+dx_{1}^{2}+dx_{2}^{2}+\frac{dz^{2}}{1-\frac{z^{4}}{z_{0}^4}}
 \Biggr\}\ .
\end{align}
To evaluate the potential between a static quark and antiquark in
this background we first need to evaluate the action of a rectangular
time-like Wilson loop whose long sides, of length ${\cal T}$, are aligned
with the $t'$ axis and whose short sides, of length $L$, are oriented
in the $x_1$ direction.  We then need to subtract the action of 
a separated quark and antiquark, each trailing a string described
as in Section 3.2.

To evaluate the Nambu-Goto action of the string worldsheet bounded
by the rectangular Wilson loop that describes the moving dipole,
we parametrize the string worldsheet by
$\tau=t'$ and $\sigma=x_{1}\in[-\frac{L}{2},\frac{L}{2}]$. The 
spacetime coordinates of the worldsheet are then given by
$(\tau, \sigma, 0, 0,z(\sigma))$, and its Nambu-Goto action (\ref{eq:ng}) is
\begin{equation}
S_{\rm NG}^{\rm dipole}
=\frac{R^{2}\cal{T}}{2\pi\alpha'}\int_{-\frac{L}{2}}^{\frac{L}{2}}d\sigma
\frac{e^{\frac{29cz^{2}}{20}}}{z^{2}}
\sqrt{\Biggl[-\sinh^{2}\eta
+\Biggl(1-\frac{z^{4}}{z_{0}^4}\Biggr)\cosh^{2}\eta\Biggr]
\Biggl(1+\frac{z'^{2}}
{1-\frac{z^{4}}{z_{0}^4}}\Biggr)}
~, \label{eq:movingdipoleaction}
\end{equation}
where we have denoted $\partial_\sigma z$ by $z'$.
We must
extremize this action in order to determine $z(\sigma)$, subject
to the boundary conditions
$z(\pm\frac{L}{2})=0$. Note that $z(\sigma)$ is symmetric about
$\sigma=0$, which is where $z(\sigma)$ reaches its maximum value
which we shall denote
$z_{\ast}$. Note also that the integration over $[-\frac{L}{2},0]$ is the
same as $[0,\frac{L}{2}]$. With a change of variables,
the action 
(\ref{eq:movingdipoleaction}) can be expressed as an integral over
$z$:
\begin{equation}
S_{\rm NG}^{\rm dipole}=\frac{R^{2}\cal{T}}{\pi\alpha'}\int_{0}^{z_{\ast}}dz\,
{\cal L}
\label{eq:movingdipoleactionZ}
\end{equation}
with the Lagrangian 
\begin{equation}
{\cal L} = \frac{e^{\frac{29cz^{2}}{20}}}{z^{2}}\sqrt{\Biggl[-\sinh^{2}\eta
+\Biggl(1-\frac{z^{4}}{z_{0}^4}\Biggr)\cosh^{2}\eta\Biggr]
\Biggl(\frac{1}{z'^2}
+\frac{1}{1-\frac{z^{4}}{z_{0}^4}}\Biggr)}
\ .
\label{eq:movingdipoleL}
\end{equation}
Since the
Lagrangian 
has no  explicit dependence on $\sigma$, the corresponding
Hamiltonian
\begin{align}
\cal{H}&=z'\frac{\partial\cal{L}}{\partial z'}-\cal{L}\nonumber\\
&=-\frac{e^{\frac{29cz^{2}}{20}}}{z^{2}}
\sqrt{\frac{\left[-\sinh^{2}\eta+\left(1
-\frac{z^{4}}{z_{0}^4}\right)\cosh^{2}\eta\right]
\left(1-\frac{z^{4}}{z_{0}^4}\right)}{1-\frac{z^{4}}{z_{0}^4}+z'^{2}}}
\label{eq:movingdipoleH}
\end{align}
is ``conserved'',  by which
we mean that it is independent of $z$. In
particular,  
\begin{equation}
\HH(z) = \HH(z_\ast) = 
-\frac{e^{\frac{29cz_{\ast}^{2}}{20}}}{z_{\ast}^{2}}\sqrt{-\sinh^{2}\eta+
\left(1-\frac{z_{\ast}^{4}}{z_{0}^4}\right)\cosh^{2}\eta}
~, \label{eq:tobesolved}
\end{equation}
where in the evaluation of $\HH(z_\ast)$ we have used the fact that
$z'=0$ at
$z=z_{\ast}$.  We can now rearrange (\ref{eq:movingdipoleH}) and
(\ref{eq:tobesolved}) to obtain an expression for $z'$, namely
\begin{equation}
z'=\pm\sqrt{\left(1-\frac{z^{4}}{z_{0}^4}\right)\left(\frac{q(z_{\ast})}{q(z)}-1\right)}\ ,
\label{eq:dzdx}
\end{equation}
where the $+$ $(-)$ sign applies for 
$-L/2 \leq\sigma\leq 0$ (for $0 \leq\sigma\leq  L/2$) and
where we have defined
\begin{equation}
q(z)\equiv\frac{e^{-\frac{29cz^{2}}{10}}z^{4}z_{0}^{4}}{z_{0}^{4}-z^{4}\cosh^{2}\eta}
\ . \label{eq:q}
\end{equation}
Upon substituting (\ref{eq:dzdx}) into (\ref{eq:movingdipoleactionZ}),
we find
\begin{equation}
S_{\rm NG}^{\rm dipole}
=\frac{\sqrt{\lambda} {\cal T}}{\pi }\int_{0}^{z_{\ast}}dz\frac{e^{\frac{29cz^{2}}{20}}}{z^{2}}
\sqrt{\frac{\left(1-\frac{z^{4}}{z_{0}^4}\right)\cosh^{2}\eta-\sinh^{2}\eta}
{\left(1-\frac{z^{4}}{z_{0}^4}\right)\left(1-\frac{q(z)}{q(z_{\ast})}\right)}}\ ,
\label{eq:dipole}
\end{equation}
where we have used $R^2/\alpha'=\sqrt{\lambda}$.

The action (\ref{eq:dipole}) 
is written in terms of $z_\ast$, the turning point of the string worldsheet,
rather than in terms of $L$, the  separation between the quark and antiquark.  $L$
and $z_\ast$ are related by
\begin{align}
\frac{L}2&= \int_{0}^{z_{\ast}}\frac{dz}{z'} \nonumber \\
&= \int_{0}^{z_{\ast}}\frac{dz}{\sqrt{\left(1-\frac{z^{4}}{z_{0}^4}\right)\left(\frac{q(z_{\ast})}{q(z)}-1\right)}}
\ . \label{eq:ell}
\end{align}
We will express our results in terms of $L$

The action (\ref{eq:dipole})  contains not only the potential between the quark-antiquark
pair but also the (infinite) masses of the quark and antiquark considered separately
in the moving medium.  
We must therefore subtract the (infinite) action $2 S^{0}_{\rm NG}$ of two
independent quarks, namely
\be
E(L) {\cal T} = S_{\rm NG}^{\rm dipole} - 2 S_{\rm NG}^{0}\ ,
\label{eq:staticpotential}
\ee
in order to extract the potential $E(L)$.
The string configuration corresponding to a single quark at rest in
the moving medium is obtained from the trailing string solution
described in our analysis of heavy quark drag in Section 3.2
by substituting (\ref{eq:xieom}) and (\ref{eq:pixifixed}) into (\ref{eq:dfaction})
and (\ref{eq:dfactionansatz}), changing variables from $r$ to $z$, and
boosting to the frame in which the quark is at rest and the plasma
is moving. We find
\be
S_{\rm NG}^{0} = 
\frac{\sqrt{\lambda} {\cal T}}{2\pi}\int_{0}^{z_{0}}dz
\frac{e^{\frac{29\,c\,z^2}{10}}}{z^2}\,
\sqrt{\frac{z^4 \cosh^2\eta - z_0^4}
{e^{\frac{29\,c\,z^2}{10}}\left( z^4 - {z_{0}}^4 \right)  
+e^{\frac{29\,c\,{z_{0}}^2}{10\cosh\eta}}\,z^4 \sinh^2\eta } }\ .
\label{eq:single}
\ee
Finally, the quark-antiquark potential $E(L)$
is obtained by substituting (\ref{eq:single})
and (\ref{eq:dipole}) into (\ref{eq:staticpotential}) and using (\ref{eq:ell}) to
relate $z_\ast$ to $L$.  We have checked that for $c=0$ these expressions all
reduce to those in Ref.~\cite{Liu:2006he}.

\FIGURE[t]{
\includegraphics[width=14cm,angle=0]{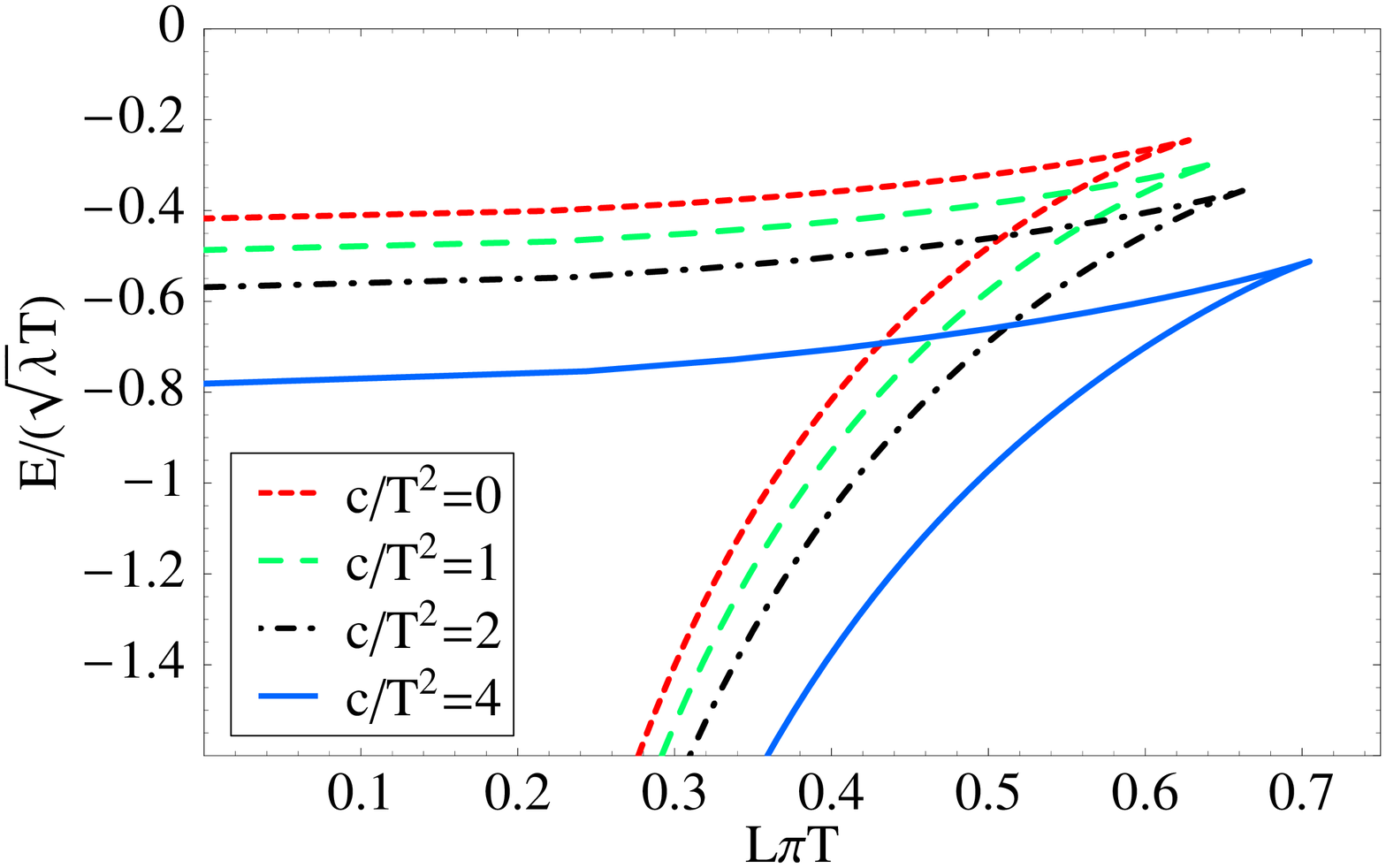}
\caption{The potential $E(L)$ between a quark and antiquark
moving through the plasma with rapidity $\eta=1$,
for four different values of the nonconformality $c/T^2$.
We plot $E/(\sqrt{\lambda}T)$ versus $L\pi T$.
Each curve has two branches that meet at a cusp at $L=L_s$,
with the lower branch being the potential of interest.
For each curve, the maximum possible $L$ at which
a string worldsheet connecting the quark and antiquark
can be found occurs at the cusp, $L=L_s$.
} \label{fig:Eal} }

In order to make a plot of $E(L)$, we 
use (\ref{eq:dipole}) and (\ref{eq:ell}) to evaluate $E$ and $L$
at a series of values of the parameter $z_\ast$,  performing the integrals numerically.
Then, in Fig.~\ref{fig:Eal} we  plot $E/(\sqrt{\lambda}T)$ versus $L\pi T$
for four values  of the nonconformality $c/T^2$, for a quark-antiquark pair
moving with rapidity $\eta=1$. Each curve has two branches that
meet at a cusp, with the cusp occurring at $L=L_s$, the largest value
of $L$ at which a string worldsheet connecting the quark and antiquark
can be found.   The lower branch is the potential of interest. The upper
branch describes unstable string configurations~\cite{Friess:2006rk}.  Two branches
arise because  $L(z_\ast)$ in (\ref{eq:ell}) is not monotonic: 
every value
of $L<L_s$ is obtained at two different values of $z_\ast$.
For $L>L_s$, $E/\sqrt{\lambda}$ vanishes. We therefore identify
$L_s$ as the screening length. (At low velocities 
this introduces a small imprecision since $E(L_s)$ is just positive and
the screening length should then be identified as the $L$ at which  
the lower branch crosses $E=0$.)

\subsection{Robustness and Infrared Sensitivity of the Screening Length}

\FIGURE[t]{
\includegraphics[width=14cm,angle=0]{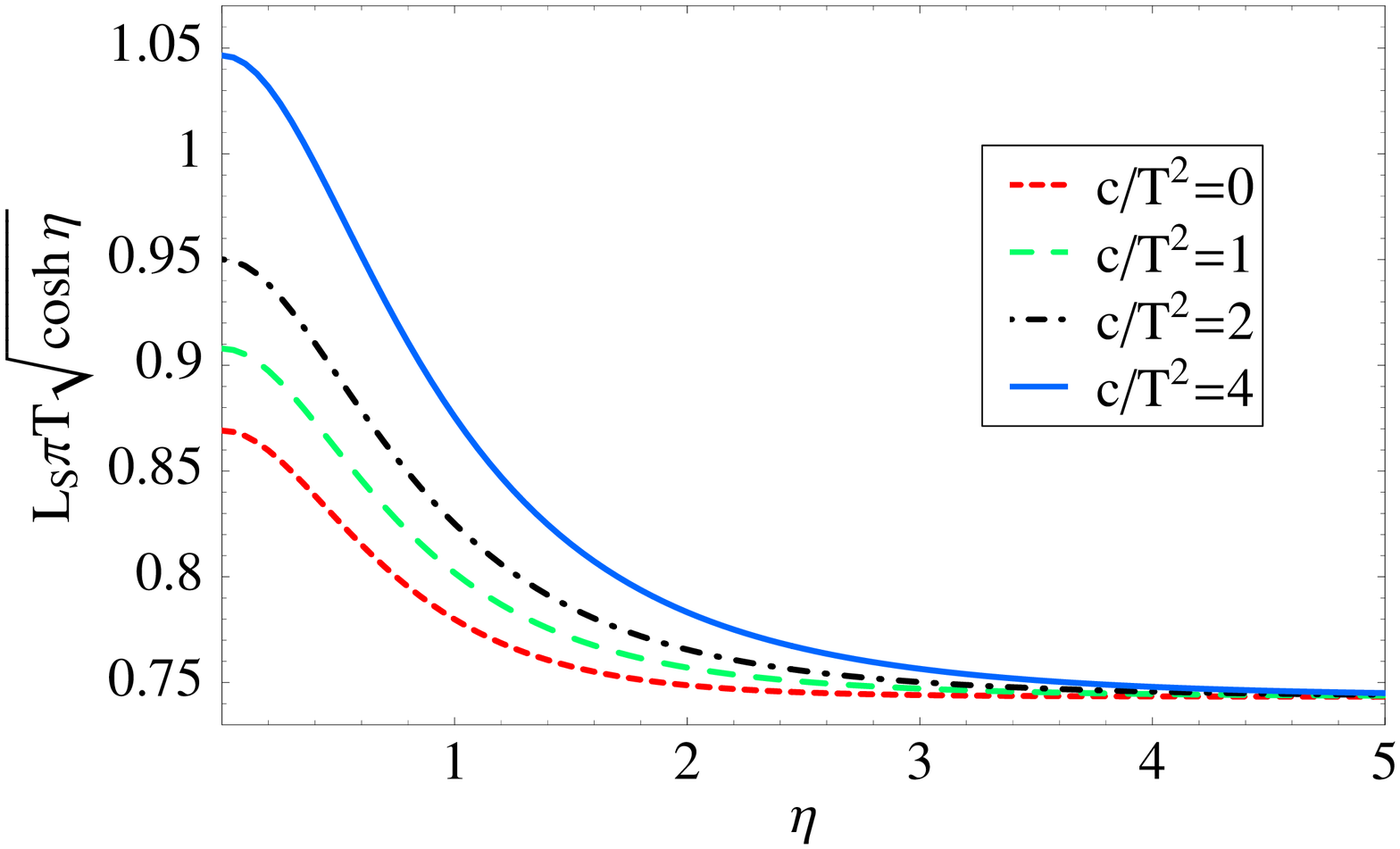}
\caption{$L_s \pi T \sqrt{\cosh\eta}$ versus rapidity $\eta$ at
four values of $c/T^2$.
}
\label{fig:sl} }

In Fig.~\ref{fig:sl}, we illustrate the velocity dependence of the
screening length $L_s$ for four values of the nonconformality $c/T^2$.
We have plotted $L_s \pi T \sqrt{\cosh\eta}$;  to the degree that the curves
are flat, we can conclude that the velocity dependence
is $L_s \pi T \propto  1/\sqrt{\cosh\eta}=(1-v^2)^{1/4}$ as in (\ref{lmax}).
We see from the figure that this is the leading velocity dependence
at large $\eta$, as can also be demonstrated analytically~\cite{Liu:2006he}.
And, we see that this leading dependence describes the velocity
dependence to within corrections of order 20\% all the way down
to $v=0$.   These conclusions hold for $c\neq 0$ as for $c = 0$,
although the corrections at small velocity grow somewhat, 
meaning that we have successfully tested the robustness of
the velocity scaling (\ref{lmax}) against the introduction
of nonconformality via $c/T^2$.  

If we now look at the effects of $c/T^2$ on the value of $L_s$,
not just on its leading velocity dependence, we see that
turning on the nonconformality  parameter results in a modest
increase in $L_s$.  The effect is greatest at $v=0$, but
even there  $L_s$
increases by only about 20\% for $c/T^2=4$.  
This means that, among the five observables that we have
analyzed and within the model we have employed,
$L_s$ is the most robust against the introduction
of nonconformality.
At large velocities,
$L_s$ becomes completely $c$-independent, meaning that at large velocities
it is infrared insensitive.  This can be understood as follows.
In order for the right-hand side of (\ref{eq:tobesolved}) to be real,
the turning point of the string worldsheet $z_\ast$, and thus
the entire worldsheet, must lie somewhere within
$0\leq z  \leq z_0/\sqrt{\cosh\eta}=z_0 (1-v^2)^{1/4}$.  This means that 
in the high velocity limit, the string worldsheet only probes
the the small-$z$, ultraviolet, region of the metric where the
effects of $c$ are not felt.  To put it more simply, since as $v\rightarrow  1$
the screening length shrinks, $L_s\pi T\propto (1-v^2)^{1/4}$, the
quark-antiquark dipole becomes sensitive only to more and more
ultraviolet physics of the plasma.

The authors of Ref.~\cite{Caceres:2006ta}
have shown that in the cascading gauge
theory of Refs.~\cite{Klebanov:2000hb,Buchel:2001gw} $L_s$ is affected by the introduction
of nonconformality even at large velocity.  This does not
contradict our conclusion that $L_s$ becomes infrared insensitive at
high velocity because this theory includes nonconformality
at all scales, not just in the 
infrared.\footnote{Our results may provide a counterexample
to a conjecture made in Refs.~\cite{Caceres:2006ta,Natsuume:2007vc}.   
Upon writing $L_s\propto (1-v^2)^p$, these authors
suggested the relationship $(\frac{1}{4}-p) \propto (\frac{1}{3}-v_s^2)$,
where $v_s$ is the velocity of sound.  We find $p=\frac{1}{4}$, but
$v_s^2$ is almost certainly not $\frac{1}{3}$ with $c\neq 0$.   
Firm conclusions cannot be drawn, however,
since, as explained in Ref.~\cite{Kajantie:2006hv} 
and Section 1, we cannot compute thermodynamic quantities like $v_s$ reliably
in the model we are employing because the deformed metric (\ref{eq:fmetric}) 
is not a solution
to supergravity equations of motion.}
Furthermore, the meson dispersion relations
analyzed in Refs.~\cite{Mateos:2007vn,Ejaz:2007hg} indicate that the size of the largest stable
mesons moving through the plasma with a given velocity shrinks
with increasing velocity in the same way that $L_s$ shrinks,\footnote{The
mesons 
are described by fluctuations of a D7-brane. 
Stable mesons moving through the plasma with a given velocity $v$ can
be found for a range of quark masses $M$ extending upward from some minimum
possible $M/T$. 
The fluctuations
corresponding to stable mesons 
with the smallest possible $M/T$ 
for a given $v$ turn out to be well localized in $z$ at the 
value of $z$ corresponding to the point where the D7-brane gets
closest to the black hole.   
According to the standard holographic relationship between position in $z$
and scale in the gauge theory,
the location in $z$ of this ``tip'' of the D7-brane therefore corresponds to
the size in the gauge theory of the largest stable mesons 
with a given propagation velocity $v$.  
The results of Ref.~\cite{Mateos:2007vn,Ejaz:2007hg} show that
this size decreases with increasing velocity proportional
to $(1-v^2)^{1/4}$, just like the screening length $L_s$.
} 
indicating
that if the meson dispersion relations were to be studied in a nonconformal
model like the one that we have analyzed, they too would become infrared 
insensitive for mesons moving with high velocity.

We have seen that
in addition to being the leading velocity dependence of $L_s$
for $v\rightarrow 1$, the expression 
 $L_s\pi T \propto (1-v^2)^{1/4}$ provides a reasonable description
 at all velocities.  This velocity dependence 
can have consequences
for the $p_T$-dependence of charmonium (bottomonium) production
in heavy ion collisions at RHIC (LHC), as it suggests
that if temperatures close to but below
that at which a particular quarkonium species dissociates
at rest are achieved, the production of this species would
drop for $p_T$ above some 
threshold~\cite{Liu:2006nn,Liu:2006he}.  
In this context, the quarkonium 
velocities that are relevant will 
not be particularly close to $v\rightarrow 1$,
meaning that $L_s$ in the relevant regime will 
not be infrared insensitive.

\section{Outlook}

We have found that the drag and momentum diffusion constants that describe
a heavy quark moving through the strongly coupled plasma and
the screening length for a quark-antiquark pair moving through
the plasma all become infrared insensitive as $v\rightarrow 1$.  Although
we used a particular toy model to diagnose this fact, in each
case we can understand it as a consequence of intrinsic attributes
of the quantity in question, meaning that the conclusion of infrared
insensitivity at high velocity transcends the particular model.
In the case of the screening length, at high velocity it becomes small
which means that in the $v\rightarrow 1$ limit 
it only probes the ultraviolet physics of the plasma.  (In the regime
of velocities accessible to quarkonium mesons produced in heavy
ion collisions, the screening length retains some infrared sensitivity.)
In the case of the drag and momentum diffusion constants, at high
velocity they are determined in their dual gravity description by the
shape and fluctuations of that portion of their trailing string worldsheet
that is outside, namely to the ultraviolet of, a worldsheet horizon that
itself moves farther and farther 
into the ultraviolet as the quark velocity increases.
In the limit of high velocity, all four of these quantities are only
sensitive to the short distance physics of the plasma, namely
to physics in a regime where the ${\cal N}=4$ SYM plasma is strongly coupled
but the quark-gluon plasma in QCD is not.
The jet quenching parameter, on the other hand, is infrared sensitive
even though it is defined at $v = 1$.  
Again, this arises from an intrinsic attribute of the quantity in question,
in this case the fact that in its dual gravity description the jet quenching
parameter is defined by a string worldsheet that extends all the
way from the ultraviolet boundary of the metric at $z=0$ to the black hole
horizon and thus probes physics of the plasma at all scales down
to of order the temperature.   

Our investigation of their infrared sensitivity provides a new illustration
of the qualitative distinction between 
the momentum diffusion constants $\kappa_T$
and $\kappa_L$ on the one hand and the jet quenching parameter $\hat q$
on the other, which arise when two noncommuting limits are taken
in opposite orders~\cite{Liu:2006he,Liu:2007ab}.   We have 
already noted that $\kappa_T$
and $\kappa_L$ are only well-defined at $v\rightarrow 1$ if 
we take this limit
while satisfying the criterion (\ref{HeavinessCriterion}), 
for example by taking
the $M\rightarrow \infty $ limit first.  If
the $M\rightarrow\infty$ limit is taken 
before the $v\rightarrow 1$ limit 
(more generally, if (\ref{HeavinessCriterion})
is satisfied), the quark trajectories for which $\kappa_T$ and $\kappa_L$ are defined
are always timelike, even as $v\rightarrow 1$.
On the
other hand, $\hat q$ is determined by a strictly light-like Wilson loop. (The light-like
Wilson lines should be thought of as describing trajectories of the gluons radiated
from the hard parton that is losing energy as it traverses the medium.)  
We can introduce
a quark mass $M$ as an ultraviolet regulator in the definition of the Wilson loop.
Then, in order to define a light-like Wilson loop in the gauge theory we must 
first take $v\rightarrow 1$, and only then take the regulator $M\rightarrow\infty$, since
if we took the limits in the opposite order the Wilson loop would not be light-like.
Our investigations show that $\hat q$, defined via the light-like
Wilson loop, is infrared sensitive: it
probes 
the properties of the strongly coupled plasma at all scales down to those
of order the temperature. In contrast, if one tries to push
$\kappa_T$ and $\kappa_L$ to $v\rightarrow 1$ while satisfying (\ref{HeavinessCriterion})
one obtains observables that 
are only sensitive
to the ultraviolet physics of the plasma, where ${\cal N}=4$ SYM is unlikely to be
a good guide to QCD.

It would be a significant advance to find other ratios of observables that
are (even close to) as universal as $\eta/s$, which is the same for all gauge theories
with dual gravity descriptions in the strong coupling and large-$N_c$ limit.
Finding infrared sensitive observables  is a prerequisite, since no infrared
insensitive observable can be universal.    Both $\eta$ and $s$ are
infrared sensitive quantities; their ratio turns out to be universal.
Our results  suggest two further infrared sensitive observables:
the jet quenching parameter $\hat q$, defined at $v=1$, and
the $v=0$ drag coefficient and momentum diffusion constant,
which are related by (\ref{Einstein}).  It is an open question whether
there are ratios involving either of these observables that are universal.

If we take as a benchmark value $c/T^2=4$, which corresponds to about
twice the level of nonconformality indicated by lattice QCD calculations
of the conformal anomaly $\varepsilon - 3 P$ at $T\sim 300$~MeV, we find
that turning on this level of nonconformality 
in the model spacetime (\ref{eq:fmetric}) that we have analyzed
increases the jet quenching
parameter by about 30\%, increases the quark-antiquark screening length
by about 20\% at low velocity, and
increases the heavy quark drag, transverse
momentum diffusion, and longitudinal momentum diffusion all
by about 80\%, again at low velocity.
The effects 
of nonconformality on the latter four
quantities all vanish at high velocities, as discussed above.  
The possibility of a significant
enhancement in the transverse momentum diffusion constant
at low velocity 
introduced by turning on a degree of nonconformality
comparable to that in QCD thermodynamics 
should be taken into account in future comparisons to charm
quark energy loss and azimuthal anisotropy as in
Refs.~\cite{Moore:2004tg,Adare:2006nq}.
Note also that the drag coefficient is no longer a 
velocity-independent constant when
nonconformality is turned on, decreasing by almost a factor of two
as $v$ is increased from near zero to near one with $c/T^2=4$.
The fact that the slowing of a moving heavy quark is no longer
governed simply by $dp/dt \propto - p$ in a strongly 
coupled but nonconformal plasma generalizes beyond the toy model
context within which we have discerned it.

Our evaluation
of the robustness of the five quantities 
we have computed against the introduction 
of nonconformality
can serve as a partial and qualitative guide to estimating 
how these quantities change in going from ${\cal N}=4$ SYM to QCD. 
A more complete understanding requires studying the effects of changing
the number of degrees of freedom in addition to introducing nonconformality.
And, our results for robustness are only quantitatively valid within the model
in which we have obtained them, making it important
to perform analyses like 
ours 
in other contexts in which nonconformality can be turned on.
Both these lines of thought serve as strong motivation for carrying out a study
like the one in this paper for the plasma of strongly
coupled  ${\cal N}=2^\ast$ gauge theory~\cite{Buchel:2003ah}.

\medskip
\section*{Acknowledgments}
\medskip

We acknowledge helpful conversations with Christiana Athanasiou,
Jorge Casalderrey-Solana,
Qudsia Ejaz, Tom Faulkner, Steven Gubser, David Mateos, Eiji Nakano,
Makoto Natsuume, Derek Teaney and 
Urs Wiedemann. 
HL is supported in part by the A.~P.~Sloan Foundation and the U.S.
Department of Energy (DOE) OJI program.  YS 
is supported in part by the MIT Undergraduate
Research Opportunity Program. This research was
supported in part by the DOE Offices of Nuclear and High Energy
Physics under grants \#DE-FG02-94ER40818 and \#DE-FG02-05ER41360.

\appendix

\section{Some Technical Details} \label{ap:A}

\subsection{Technical Details Needed in the Calculation of $\kappa_T$}

In Section 3.4, we calculate the 
transverse momentum diffusion constant $\kappa_T$ by determining
the two point function for the transverse fluctuations of
the worldsheet of the trailing string.  The equation of motion for
the
Fourier transform of the transverse fluctuations, $Y_\omega$,
is given in (\ref{eq:eleqtransfourier}), and solutions
with the correct behavior (\ref{eq:factorout}) near the
horizon $u=1$ are then specified by the ordinary differential
equation (\ref{eq:odeforF}) for $F(\omega,u)$, defined
in (\ref{eq:factorout}).  The differential equation (\ref{eq:odeforF}) 
contains
two functions $X$ and $V$ that we did not specify in Section 3.4.
To lowest  order in $\omega$, these functions are given by 
\begin{equation}
X=X_{1}\omega+\OO(\omega^{2}) ~, \label{eq:TXe}
\end{equation}
and
\begin{equation}
V=V_{0}+\OO(\omega) ~, \label{eq:TVe}
\end{equation}
where
\begin{align}
X_{1}=& -\frac{i\,\sqrt{\gamma}}{20\pi \,
    {\left( 1 - u^2 \right) }^2\,{\sqrt{4\,{\pi }^2 - \frac{29\,c\,v^2\,{\sqrt{1 - v^2}}}{5\,T^2}}}}
\Biggl\{ 
-40\,{\pi }^2u^2 - 20\,{\pi }^2\left( 1 - u^2 \right) \nonumber \\
&\qquad \qquad +\frac{ 1 }{T^2 
         \left[ - u^2\,v^2 +
           e^{-\frac{29\,c\,\left( 1 - u \right) \,{\sqrt{1 - v^2}}}{10\,{\pi }^2\,T^2}}\,
            \left( 1 - u^2\,\left( 1 - v^2 \right)  \right)  \right] } 
\Biggl[ 10\,{\pi }^2T^2 u^2\left( 1 - 3 u^2 \right) v^2\nonumber\\
&\qquad \qquad \qquad \qquad 
-e^{-\frac{29\,c\,\left( 1 - u \right) \,{\sqrt{1 -
v^2}}}{10\,{\pi }^2\,T^2}}\,
            \Bigl( 29\,c\,u ( 1 - u^2 )
               ( 1 - u^2 ( 1 - v^2 )  ) \sqrt{1-v^2} \nonumber \\
&\qquad \qquad \qquad \qquad \qquad \qquad -
              10\,{\pi }^2 T^2 ( 1 + u^2 ( 2 - v^2 )  -
                 3\,u^4 ( 1 - v^2 ) ) \Bigr)   \Biggr] \Biggr\}
\label{eq:X1}
\end{align}
and
\begin{align}
V_{0}=& -\frac{1}{20\,{\pi }^2 T^2 u \left( 1 - u^2 \right) 
    \Bigr( - u^2 v^2   + e^
        {-\frac{29\,c\,\left( 1 - u \right) \,{\sqrt{1 - v^2}}}{10\,{\pi }^2\,T^2}}\,
       \left( 1 - u^2\,\left( 1 - v^2 \right)  \right) \Bigr) } 
\Biggl\{ 10\,{\pi }^2 T^2 u^2 \left( 1 - 3 u^2 \right) v^2 \nonumber\\
&\qquad \qquad -
    e^{-\frac{29 c \left( 1 - u \right) \,{\sqrt{1 - v^2}}}{10\,{\pi }^2\,T^2}}\,
     \Bigl[ 29\,c u ( 1 - u^2 ) 
        \left( 1 - u^2 ( 1 - v^2 )  \right) \sqrt{1-v^2} \nonumber \\
& \qquad\qquad\qquad\qquad\qquad\qquad
-       10\,{\pi }^2 T^2 \left( 1 + u^2 ( 2 - v^2 )  - 3\,u^4 ( 1
- v^2 )
            \right) \Bigr]\Biggr\}\ .
\label{eq:Y0}
\end{align}
In this Appendix, we shall solve the equation (\ref{eq:odeforF})
for $F$, first to zeroth order in $\omega$ and then to first order.

To zeroth order in $\omega$, (\ref{eq:odeforF}) reads
\begin{equation}
V_{0}\partial_{u}F+\partial_{u}^{2}F=0 ~.
\label{eq:zerothorderode}
\end{equation}
We can solve this equation upon noticing that
near $u=1$, $V_{0}$ can be expanded in powers of $(1-u)$ 
and takes on the simple form
\begin{equation}
V_{0}=-\frac{1}{1 - u}+\OO(1) ~. \label{eq:vnotexp}
\end{equation}
This means that the only solutions to
(\ref{eq:zerothorderode}) that are regular at $u=1$
are constant solutions, with $\partial_u F=0$.  Normalizing
$Y_\omega$ such that $Y_\omega\rightarrow 1$ in the $u\rightarrow 0$
limit corresponds to choosing the constant solution $F=1$.
This normalization is required if one is to preserve the 
standard AdS/CFT
relationship between the fluctuations of the 
string worldsheet in the bulk, $\delta y$, and operators
and sources in the gauge theory on the boundary at $u=0$. In
particular, it is required in order for 
the retarded propagator $G_R$ to be given
by (\ref{eq:GRintermsofZ}).
The same normalization is also used in the calculation of
$\kappa_{L}$.

Now, working to first order in $\omega$ and knowing that $F=1$
to zeroth order, we write $F$ as
\begin{equation}
F=1+\omega Z \ , \label{eq:fomegaz_App}
\end{equation}
a form that we used in (\ref{eq:fomegaz}).
The first order terms in the differential equation
(\ref{eq:odeforF}) then become
\begin{equation}
X_{1}+V_{0}\partial_{u}Z+\partial_{u}^{2}Z=0\ .
\label{eq:firsrorderode}
\end{equation}
We now define
\begin{equation}
W\equiv - i \partial_{u}Z \ ,
\label{eq:defW}
\end{equation}
and obtain a first order differential equation for $W(u)$ given by
\begin{equation}
\frac{X_{1}}{i}+V_{0}W+\partial_{u}W=0 ~, \label{eq:foodeforw}
\end{equation}
where $X_{1}$ and $V_{0}$ are given by (\ref{eq:X1}) and
(\ref{eq:Y0}).  Note that since $X_1$ is imaginary the equation
(\ref{eq:foodeforw}) for $W$ has real coefficients.  
The differential
equation (\ref{eq:foodeforw}) can be solved 
analytically, yielding
\begin{align}
W=& - \frac{\sqrt{u}\sqrt{\gamma}}
{\left( 1 - u^2 \right) \,{\sqrt{20\,{\pi }^2 -
\frac{29\,c v^2 {\sqrt{1 - v^2}}}{T^2}}}\,
      {\sqrt{e^{-\frac{29\,c {\sqrt{1 - v^2}}}{10\,{\pi }^2\,T^2}}\Bigl(
-e^{\frac{29\,c\,{\sqrt{1 - v^2}}}{10\,{\pi }^2\,T^2}}\,u^2 v^2  +
            e^{\frac{29\,c\,u\,{\sqrt{1 - v^2}}}{10\,{\pi }^2\,T^2}}\,
             \left( 1 - u^2 \left( 1 - v^2 \right)  \right) \Bigr) }}} 
\nonumber \\
&\qquad\qquad \times \Biggl\{  \pi 
 {\sqrt{5 u \, e^{-\frac{29\,c\,{\sqrt{1 - v^2}}}{10\,{\pi
}^2\,T^2}}\,
\Bigl( - e^{\frac{29\,c\,{\sqrt{1 - v^2}}}{10\,{\pi }^2\,T^2}}\,u^2 v^2  +
        e^{\frac{29\,c\,u\,{\sqrt{1 - v^2}}}{10\,{\pi }^2\,T^2}}\,
\left( 1 - u^2\left( 1 - v^2 \right)  \right)  \Bigr) }} \nonumber \\
&\qquad\qquad\qquad\qquad+C\,{\sqrt{1 - u^2}}\,{\sqrt{20\,{\pi }^2 -
\frac{29\,c\,v^2 {\sqrt{1 - v^2}}}{T^2}}}\Biggr\} \ , \label{eq:Wsolkt}
\end{align}
where the integration constant $C$ has to be determined by the
requirement that $W$ must be regular at the 
worldsheet horizon $u=1$.  
To determine $C$, we expand
(\ref{eq:Wsolkt}) 
about $u=1$, which yields
\begin{equation}
W=\frac{\sqrt{\gamma}\,\left( C +\frac{1}{2} \right) \,\pi }{{\sqrt{4\,{\pi }^2 -
\frac{29\,c\,v^2\,{\sqrt{1 - v^2}}}{5\,T^2}}}}
\frac{1}{u-1} +\OO(1) \label{eq:Weuo}
\end{equation}
The coefficient of the $1/(u-1)$ term in
(\ref{eq:Weuo}) must vanish, which determines that $C=-\frac{1}{2}$.
With $C$ determined, (\ref{eq:Wsolkt}) constitutes a fully
explicit expression for $W$, which according to (\ref{eq:defW}) 
should then be integrated to give $Z$. The further integration constant
in $Z$ is fixed by the requirement that 
$Y_\omega\rightarrow 1$, meaning that $Z\rightarrow 0$, 
for $u\rightarrow 0$.  In
our calculation of $\kappa_T$, we do not need the entire function $Z$.
According to (\ref{eq:GRintermsofZ}), 
all we need is the leading term in $Z$
(or $W$) at $u\rightarrow 0$.  From (\ref{eq:Wsolkt}) we determine
that $W\propto \sqrt{u}$ in the $u\rightarrow 0$ limit, and upon
integrating to determine $Z$ in this regime we obtain (\ref{z3}).

\subsection{Technical Details Needed in the Calculation of $\kappa_L$}

The technical details needed in the calculation of $\kappa_L$
are completely analogous to those described in Appendix A.1,
including in particular the logic of how the boundary
conditions are satisfied.  The only difference is in the
functions $X_1$ and $V_0$, which in the longitudinal case
are given by
\begin{align}
X_{1}=&-\frac{i\,\sqrt{\gamma}}{20\,\pi {( 1 - u^2 ) }^2
    {\sqrt{4\,{\pi }^2 - \frac{29\,c\,v^2\,{\sqrt{1 - v^2}}}{5\,T^2}}}}
\Biggl\{-40\,{\pi }^2 u^2 - 20\,{\pi }^2 ( 1 - u^2 ) \nonumber \\
&\qquad\qquad+\frac{1}{T^2 \Bigl( - u^2 v^2   +
      e^{-\frac{29\,c\,( 1 - u ) \,{\sqrt{1 - v^2}}}{10\,{\pi }^2\,T^2}}\,
       \left( 1 - u^2\,( 1 - v^2 )  \right)  \Bigr) }\Biggl[
10 {\pi }^2 u^2 v^2 T^2 ( 5 - 3 u^2 ) \nonumber \\
&\qquad\qquad\qquad\qquad-
    58\,c u^3 v^2 ( 1 - u^2 ) \,{\sqrt{1 - v^2}}  \nonumber \\
& \qquad\qquad\qquad\qquad
-e^{-\frac{29\,c\,( 1 - u ) \,{\sqrt{1 - v^2}}}{10\,{\pi }^2\,T^2}}
\Biggl(29\,c u ( 1 - u^2 )  {\sqrt{1 - v^2}} 
\left( 1 - u^2\,( 1 - v^2 )  \right ) \nonumber\\
&\qquad\qquad\qquad\qquad\qquad\qquad 
+ 30 \pi^2 u^4 T^2 (1-v^2) 
-  10\,{\pi }^2\,T^2\,\left ( 1 + u^2\,( 2 - 5\,v^2 )\right ) \Biggr)\Biggr]
\Biggr\} 
\ ,
\label{eq:X1AGAIN}
\end{align}
and
\begin{align}
V_{0}=&-\frac{1}{20\,{\pi }^2 T^2 u ( 1 - u^2 ) 
    \Bigl( - u^2 v^2   
+ e^{-\frac{29\,c\,( 1 - u ) \,{\sqrt{1 - v^2}}}{10\,{\pi }^2\,T^2}}
       \left( 1 - u^2 ( 1 - v^2 )  \right)  \Bigr) } \nonumber \\
&\qquad\qquad\times \Biggl\{ 2 u^2 v^2 \Bigl( 5 {\pi }^2 T^2 ( 5 - 3 u^2 )  -
     29\,c u ( 1 - u^2 )  {\sqrt{1 - v^2}} \Bigr) \nonumber \\
&\qquad\qquad\qquad\qquad -
  e^{-\frac{29\,c\,( 1 - u ) \,{\sqrt{1 - v^2}}}{10\,{\pi }^2\,T^2}}\,
   \Biggl[ 29\,c u ( 1 - u^2 ) {\sqrt{1 - v^2}} 
\left( 1 - u^2 ( 1 - v^2 )  \right) \nonumber \\
&\qquad\qquad\qquad\qquad\qquad\qquad -
     10\,{\pi }^2 T^2 \Bigl( 1 + u^2 ( 2 - 5 v^2 )  - 3 u^4 ( 1 - v^2 )  
\Bigr)  \Biggr] \Biggr\} \ .
\label{eq:Y0AGAIN}
\end{align}


\end{document}